\documentclass[12pt]{report}

\usepackage{amsxtra}
\usepackage{amsopn}
\usepackage{amsbsy}
\usepackage{amstext}
\usepackage{amsmath, amsfonts}
\usepackage{dsfont}
\usepackage{autobreak}
\usepackage{docmute}
\usepackage{dcolumn}
\usepackage{amsthm}
\usepackage{epsfig}
\usepackage{graphicx}
\usepackage{graphics}
\usepackage{latexsym}
\usepackage{rotating}
\usepackage[usenames]{color}
\usepackage{yfonts}
\usepackage{float}
\usepackage{ucs}
\usepackage{xspace} 
\usepackage{mathrsfs}
\usepackage{booktabs}
\usepackage[toc,page]{appendix}
\usepackage{array}
\usepackage[normalem]{ulem}
\usepackage{tikz}
\usepackage[super]{nth}
\usepackage[margin=2cm]{geometry}
\usepackage{xr-hyper}
\usepackage[colorlinks=true]{hyperref}
\usepackage{exercise}
\usepackage[most]{tcolorbox}
\usepackage{setspace}

\newcommand{\eff}{\text{eff}}
\newcommand{\SeffL}{\vec{S}_{\eff}\cdot \vec{L}}
\newcommand{\Eff}{\vec{S}_{\eff}\cdot \vec{L}}
\newcommand{\pb}[1]{\left\lbrace  #1   \right\rbrace  }
\newcommand{\vv}[1]{ \vec{#1}}
\newcommand{\mc}[1]{ \mathcal{#1}}
\newcommand{\cJ}{\mathcal{J}}
\newcommand{\pd}{\partial}

\allowdisplaybreaks

\newtcolorbox[auto counter,number within=chapter]{definition}[1][]{
  enhanced,
  breakable,
  fonttitle=\scshape,
  title={Box \thetcbcounter},
  #1
}

\begin{document}

\begin{titlepage}

\begin{center}
{\Large Integrability and action-angle-based solution of the post-Newtonian BBH system}\\
(LECTURE NOTES)\\
\end{center}
\begin{center}
Sashwat Tanay (sashwattanay@gmail.com)
\linebreak
\linebreak
\linebreak
\linebreak
\linebreak
For the most updated version of these lecture notes (with typos fixed), 
see the GitHub version \href{https://github.com/sashwattanay/lectures_integrability_action-angles_PN_BBH/blob/gh-action-result/pdflatex/lecture_notes/main.pdf}{here}.
To cite this article, please use the arXiv version 
\href{https://arxiv.org/abs/2206.05799}{here}.
\end{center}
\newpage
\begin{center}
{\Large ACKNOWLEDGMENTS}\linebreak
\linebreak
\end{center}
I would like to thank
Nicol\'as Yunes for an invitation to conduct a lecture workshop 
at the University of Illinois Urbana-Champaign, for
these lecture notes were initially prepared for the workshop.
I am also grateful to Rickmoy Samanta
and Jos\'e T.~G\'alvez Ghersi
for carefully reading the manuscript
and providing useful comments.
\end{titlepage}

\tableofcontents

\chapter{Introduction}

\section{The lecture notes}

These lecture notes are based on Refs.~\cite{tanay2021action, tanay2021integrability, Cho:2019brd}
 which aim to give closed-form solutions to the spinning, eccentric
binary black hole dynamics at 1.5PN via two different equivalent ways: (i)
the standard way of integrating Hamilton's equations and (ii) using action-angle variables.

The above papers assume a certain level of familiarity
with the symplectic geometric approach to classical mechanics, the non-fulfillment
of which on the reader's part may make the papers appear esoteric.
The purpose of these lecture notes is to give the reader this prerequisite
 knowledge which the above papers assume on the reader's part. 

Although these notes are meant to be pedagogical 
in the exposition, but they
lack rigor. If the reader 
finds these notes to be incomplete or lacking rigor, they are welcome to
refer to the sources cited in these notes as well as the above papers.

We use two kinds of filled boxes in these lecture notes
\begin{itemize}
\item The boxes with an explicit label ``Box'': These are meant
to give the reader a reference to more advanced sources to supplement
the material discussed herein. 
\item The boxes without an explicit label ``Box'': These simply 
summarize the important points.
\end{itemize}

\section{The \textsc{Mathematica} package}

Accompanying the above three papers (Refs.~\cite{tanay2021action, tanay2021integrability, Cho:2019brd}), and these lecture notes
is a \textsc{Mathematica} package \cite{MMA1}. This package can 
do the following
\begin{itemize}
\item can perform numerical integration of the 1.5PN equations of motion (EOMs).
\item can implement the analytical solution presented in Ref.~\cite{Cho:2019brd} with
1PN effects included.
\item can implement the action-angle-based analytical solution presented in
Refs.~\cite{tanay2021action, tanay2021integrability}
\item can compute the frequencies (within the action-angles framework) 
of the 1.5PN spinning BBHs
\item can compute the Poisson bracket between any two quantities.
\end{itemize}

\textbf{How to run the \textsc{Mathematica} package?} 
See the YouTube video \href{https://youtu.be/aoiCk5TtmvE}{here}
which demonstrates how to use the package.

\chapter{The setup}

\section{Statement of the problem}     \label{statement}

\begin{figure}
 \centering
  \includegraphics[width=0.9\linewidth]{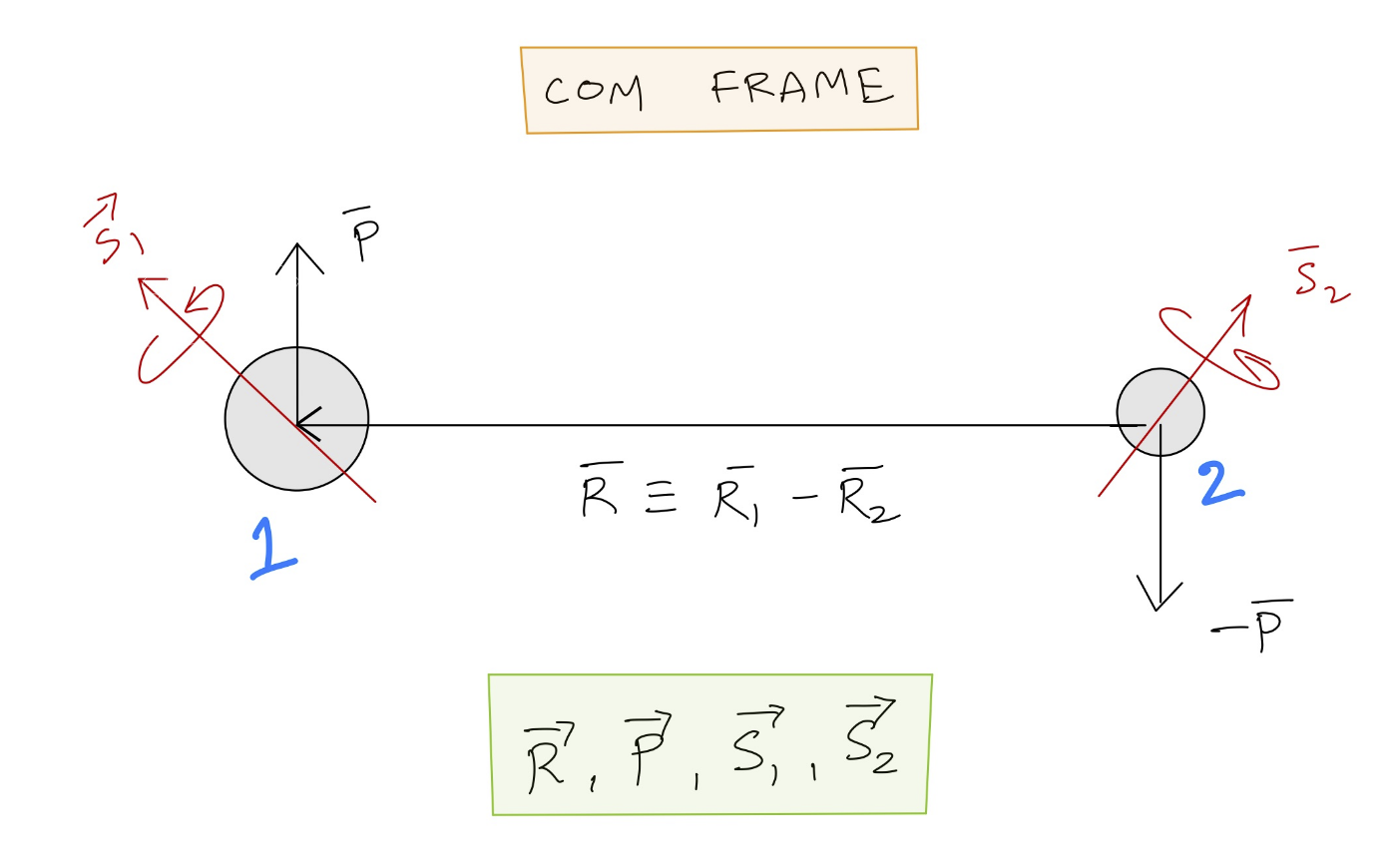}
  \caption{Schematic setup of a precessing black hole
    binary. All phase space variables are contained in the
    form of four 3D vectors $\vv{R}, \vv{P}, \vv{S}_1$ and $\vv{S}_2$.
    \vspace{-1.em}
  }
  \label{fig:SetupFig}
\end{figure}

We start by describing the canonical variables and the dynamical
setup used to study eccentric binaries of black holes with precessing
spins in the post-Newtonian (PN) approximation. 
The BBH system under consideration is
schematically displayed in Fig.~\ref{fig:SetupFig}, using its
center-of-mass frame \cite{Damour:1988mr}, to define the separation
vector $\vec{R} \equiv \vec{R}_1 -\vec {R}_2$ and the linear momenta
$\vec{P} \equiv \vec{P}_1 = - \vec{P}_2$ of a binary of black holes
with masses $m_1$ and $m_2$. With these quantities, we build the
Newtonian orbital angular momentum $ \vec{L} \equiv \vec{R} \times \vec{P }$,
and the total angular momentum
$\vec{J} \equiv \vec{L} + \vec{S}_{1}+ \vec{S}_{2}$ which includes the BH
spins $\vec{S}_{1}$ and $ \vec{S}_{2}$. The individual BH masses are
$m_1$ and $m_2$ and the total mass $M \equiv m_1+m_2 $.
Additionally, the reduced mass is given by $\mu\equiv m_1m_2/M$ and the symmetric
mass ratio $\nu\equiv\mu/M$ is a function of the reduced mass.
The constants $\sigma_1 \equiv 1 + 3 m_{2}/4 m_{1}$ and
$\sigma_2 \equiv 1 + 3 m_{1}/4 m_{2}$ are used to build the effective
spin
\begin{equation}
  \vec{S}_{\mathrm{eff}} \equiv  \sigma_1 \vec{S}_{1} + \sigma_2\vec{S}_2 \,.
  \label{eq:seff}
\end{equation}

In terms of the scaled variables $\vec{r} \equiv \vec{R} / G M, \vec{p} \equiv \vec{P} / \mu$,
the 1.5 post-Newtonian (PN) Hamiltonian is given by 
\cite{Barker:1966zz, Damour:2001tu,
  Barker:1975ae, Hartl:2004xr,Steinhoff:2010zz}
\begin{align}
  H  =  H_{\mathrm{N}}   +    H_{1\mathrm{PN}}  +    H_{1.5\mathrm{PN}}   + \mathcal{O}(c^{-4})
  \,,
  \label{Hamiltonian-1}
  \end{align}
 where 
\begingroup
\allowdisplaybreaks
\begin{align}
H_{\mathrm{N}}       ={} &   \mu\left(\frac{p^2}{2}     -    \frac{1}{r}\right)  ,      \label{eq:int1a}    \\
H_{\mathrm{1PN}}     ={} &   \frac{\mu}{c^2}\bigg\{  \frac{1}{8}(3 \nu-1) p^4     +  \frac{1}{2r^2}       \nonumber  \\
   &   - \frac{1}{2 r}  \left[   (3+ \nu) p^2  +  \nu (\hat{r}\cdot \vec{p})^2  \right]\bigg\}    ,        \\
H_{\mathrm{1.5PN}}   ={} &   \frac{ 2 G }{c^{2} R^{3}} \SeffL   ,             \displaybreak[0]
   \,.
\label{Hamiltonian-2}
\end{align}
\endgroup

\textbf{The problem:} The above 1.5PN Hamiltonian is  a function of the 
phase-space variables $\vec{R}(t), \vec{P}(t), \vec{S}_1(t)$ and $\vec{S}_2(t) $
only. Our challenge is to integrate the Hamilton's equations
obtained from this Hamiltonian to obtain the solution $\vec{R}(t),
 \vec{P}(t), \vec{S}_1(t)$ and $\vec{S}_2(t) $. This is the subject of these 
 lecture notes.

\section{Defining the 1.5PN BBH system}        \label{defining the system}

In graduate level classical mechanics, specification of the Hamiltonian 
$H( \vv{p}, \vv{q})$ is considered equivalent to specifying the system, for 
the application of Hamiltons's equations  \cite{goldstein2013classical}
\begin{align}    \label{Hamilton's_eqns}
\dot{q_i} =  \frac{\partial H}{\partial p_i},~  ~~~~~~~~~~~~~
\dot{p_i} = - \frac{\partial H}{\partial q_i},
\end{align}
immediately furnishes the equations 
of motion (EOMs). Eqs.~\eqref{Hamilton's_eqns} further imply  that
any function $f(p,q)$ of the phase space variables obeys (Eq. 9.94 of
Ref.~\cite{goldstein2013classical})
\begin{align}    
\frac{d f}{d t}   & = \pb{f, H} + \frac{\partial f}{\partial t} = \pb{f, H}, ~~~~~~~~~~~~~~~~\text{where}    \label{EOM-PB}  \\
\{f, g\} &\equiv  \sum_{i=1}^{N}\left(\frac{\partial f}{\partial q_{i}} \frac{\partial g}{\partial p_{i}}-\frac{\partial f}{\partial p_{i}} \frac{\partial g}{\partial q_{i}}\right)    \label{PB_defined}
\end{align}
where the last equality in Eq.~\eqref{EOM-PB}
 is valid if the Hamiltonian $H$ is not a 
function of time (which will be the case for our BBH system, the subject of
these lecture notes). $\pb{f,g}$ is called the Poisson bracket (PB)
between $f$ and $g$. Actually,
\begin{center}
Eq.~\eqref{Hamilton's_eqns}    $\Leftrightarrow$   Eq.~\eqref{EOM-PB},
\end{center}
i.e., the implication goes both ways; the two equations are equivalent.

Our point of view towards the PB-based EOMs for these lecture notes will be somewhat 
different from the above standard approach adopted by classical mechanics texts. 
For the BBH system with 
phase-space variables $\vec{R}(t), \vec{P}(t), \vec{S}_1(t)$ and $\vec{S}_2(t) $,
the EOMs for any general function 
$f(\vec{R}(t), \vec{P}(t), \vec{S}_1(t),\vec{S}_2(t))$ is still given by
Eq.~\eqref{EOM-PB}.   
But instead of Eq.~\eqref{PB_defined}, we define the PBs via
\begin{align}     \label{PBs_defined_1}
\left\{R_{i}, P_{j}\right\}=\delta_{{j}{i}} \quad \text { and } \quad\left\{S_{A}^{i}, S_{B}^{j}\right\}=\delta_{A B} \epsilon_{k}^{i j} S_{A}^{k}, 
\end{align}      
where the labels $A$ and $B$ refer to the two BHs. How to
define the PB between any two functions $f$ and $g$ 
of the phase-space variables? We simply axiomatize the anti-commutativity,
bilinearity, product,
and the chain rules for the PBs
\begin{subequations}     \label{PBs_defined_2}
\begin{equation}
\pb{f,g}   =  - \pb{g,f}   ,   
\end{equation}
\begin{equation}
\{a f+b g, h\}=a\{f, h\}+b\{g, h\}, \quad\{h, a f+b g\}=a\{h, f\}+b\{h, g\}, \quad a, b \in \mathbb{R}    ,  
\end{equation}
\begin{equation}
\{f g, h\}=\{f, h\} g+f\{g, h\}  ,
\end{equation}
 \begin{equation}
\pb{f, g (v_i)}  =  \pb{f, v_i}  \frac{\pd g}{\pd v_i}   ,  
\end{equation}
\end{subequations}
where $v_i$ represents any of the phase-space variables.
The PB between any two phase space variables, i.e. components of
$\vec{R}(t), \vec{P}(t), \vec{S}_1(t)$ and $\vec{S}_2(t) $ which 
does not fall under the purview of
 Eqs.~\eqref{PBs_defined_1} and \eqref{PBs_defined_2} is assumed to vanish.
 Note that Eq.~\eqref{PB_defined} implies Eqs.~\eqref{PBs_defined_2}, 
 but for these lecture notes,
  we will take the latter as definitions and totally disregard 
 the former.

We therefore have defined our system of interest completely in that 
we can now write its EOM. This definition consists of
\begin{itemize}
\item specifying the Hamiltonian via Eq.~\eqref{Hamiltonian-1}.
\item axiomatizing Eqs.~\eqref{PBs_defined_1}, and \eqref{PBs_defined_2}.
 These enable us to evaluate the PB between any 
two functions of the phase-space variables.
\item stating the EOM via Eq.~\eqref{EOM-PB}.
\end{itemize}
Note that in this slightly different point of view,
we don't try to define the PBs via partial derivatives 
as in Eq.~\eqref{PB_defined}. Also, note that it appears as if there is no way
to recover the spin PB of Eq.~\eqref{PBs_defined_1} via Eq.~\eqref{PB_defined}.
In this sense, this new way of defining the PB seems more general.

\begin{Exercise}    \label{exercise-1}
\textbf{Problem:} Compute the PB 
\begin{align}
\pb{R_x , \sin  P_x   +   P_x  }.
\end{align}

\textbf{Solution:} 
\begin{align}
& \pb{R_x , \sin  P_x   +   P_x  }  ,   \\ 
& =  \pb{R_x,  \sin  P_x  }  +  \pb{R_x,   P_x  }  ,  \\
& =  \pb{R_x,    P_x  }  \frac{\pd \sin P_x}{\pd P_x}  +  \pb{R_x,   P_x  }  ,  \\
&  =   \cos P_x + 1.
\end{align}
Use has been made of the second and the fourth of Eqs.~\eqref{PBs_defined_2}
in the above manipulations.
\end{Exercise}

We have prepared a
\textsc{Mathematica} package \cite{MMA1}
which can compute the PB between any two quantities,
based on the PB theory discussed above.

If we try to compute the PB between the azimuthal angle 
$\phi_{A} = \arctan (S_A^y/S_A^x)$ of the spin vector
of a BH and its $z$-component, it can be checked (using 
Eqs.~\eqref{PBs_defined_1} and \eqref{PBs_defined_2}) that it comes out to be
\begin{equation}
\left\{\phi_{A}, S_{B}^{z}\right\}=\delta_{A B}  ,     \label{spin-PB}
\end{equation}
which upon comparison with Eqs.~\eqref{PBs_defined_1}
makes us conclude that $\phi_A$ and $S_{A}^{z}$ respectively ``act
like'' position and momentum variables respectively. This is of big significance 
when we deal with the action-angle variables (AAVs) later. AAVs are the key
to obtaining the closed-form solutions we have set out to seek. 
We end this section 
with a definition. \\ \\
\textbf{Commuting quantities: } Two quantities {commute} if their PB vanishes.

\chapter{Integrable systems and action-angle variables}      \label{define_integrable_sys}

\section{Definitions}        \label{definition}

We will focus our attention on systems which possess a time-independent
Hamiltonian \\

\textbf{Hamiltonian systems:} A dynamical system possessing a 
Hamiltonian $H(\vv{p}, \vv{q})$ and whose EOMs are given 
via Hamilton's equations.
\begin{align}   
\dot{q_i} =  \frac{\partial H}{\partial p_i},~  ~~~~~~~~~~~~~
\dot{p_i} = - \frac{\partial H}{\partial q_i}, \\
\end{align}

\textbf{Canonical transformation:} For a Hamiltonian system with the
Hamiltonian $H(\vv{p}, \vv{q})$, a transformation $\vv{Q}(\vv{p}, \vv{q}),
\vv{P}(\vv{p}, \vv{q})$ is called canonical if Hamilton's equations in
the old coordinates imply Hamilton's equations in the new coordinates, i.e.
\begin{align}
& \dot{q_i} =  \frac{\partial H}{\partial p_i},~  ~~~~~~~~~~~~~
\dot{p_i} = - \frac{\partial H}{\partial q_i}    \\
\implies &  \dot{Q_i} =  \frac{\partial K}{\partial P_i},~  ~~~~~~~~~~~~~
\dot{P_i} = - \frac{\partial K}{\partial Q_i}, 
\end{align}
where $K (\vec{P}, \vec{Q}) =
   H(   \vv{p}(\vv{P}, \vv{Q}), \vv{q}(\vv{P}, \vv{Q}) )$. \\

\textbf{Integrable system and action-angle variables:} 
We will define integrable  systems and action-angle variables (AAVs)
in one shot.
For integrable systems, 
canonical transformation $( \vec{p}, \vec{q}) \leftrightarrow $ {$ (  \vec{\cal{J}} , \vec{\theta})$} exists such that {$H = H(\vec{\mathcal{J}})$}
(or rather $ \pd H/ \pd \theta_i = 0$)
and also
{$\{ \vec{p}, \vec{q} \}(\theta_i + 2 \pi)  = \{ \vec{p}, \vec{q} \}(\theta_i ) $}.
The first equation means that the Hamiltonian depends only on the actions 
and not the angles.
The last equation means that $\vv{p}$ and $\vv{q}$ are $2 \pi$-periodic
functions of $\theta_i$'s. $\mc{J}_i$'s and $\theta_i$'s are respectively called
the action and the angle variables.

\section{Elementary properties of action-angle variables}

The above definition of action-angle variables may seem ad-hoc 
but some very useful conclusions follow from this definition.
First note that due to 
 $( \vec{p}, \vec{q}) \leftrightarrow $ {$ (  \vec{\cal{J}} , \vec{\theta})$} 
 being a canonical transformation, the actions 
 $\vec{\cal{J}} $ act like the new momenta variable and the 
 angles $\vec{\theta}$ act like the new position variables.

 Writing Hamilton's equations in terms of these new momenta and
 positions (actions and angles), we get
\begin{align}              
\dot{\mathcal{J}}_{i}  &=   -  \frac{\partial H}{ \partial \theta_{i}}    =0    &&   \Longrightarrow \mathcal{J}_{i} \text { stay constant } ,  \label{AA_eqn_1}   \\ 
\dot{\theta}_{i}   &=   \frac{\partial H }{ \partial \mathcal{J}_{i}} \equiv \omega_{i}(\overrightarrow{\mathcal{J}})     && \Longrightarrow \theta_{i}=\omega_{i}(\overrightarrow{\mathcal{J}}) t. \label{AA_eqn_2} 
\end{align}
We have chosen to 
call $\partial H/ \partial \mathcal{J}_{i} $ the frequencies $\omega_i$'s
since they denote the linear rate of increase of the 
corresponding angles $\theta_i$'s. These frequencies $\omega_i$'s are 
constants too since they are functions of only the constant $\mc{J}_i$'s.

More interesting and useful conclusions follow. It's clear that we know what 
actions and angles are at any later time given their values at an initial time
(actions stay constants, angle increase linearly with time at a known rate).
Therefore, \textit{if we know how to switch back} from $(  \vec{\cal{J}} , \vec{\theta})$
to $(\vv{p}, \vv{q})$, then we can have $\vv{p}(t), \vv{q}(t)$ for any later time $t$,
that is to say, we can have the solution of the system.

We now summarize the above conclusions.\\
\begin{tcolorbox}
From the above definition of action-angle variables, it follows that
\begin{itemize}
\item Actions are constants.
\item Angles increase linearly with time at constant  rate $\omega_i  =  \partial H/ \partial \mathcal{J}_{i} $ .
\item Having action-angle variables can help us have the solution $\vec{p}(t), \vec{q}(t)$ of the system.
\end{itemize}
\end{tcolorbox}

\section{Liouville-Arnold theorem}

\subsection{Statement of the theorem}

Stated loosely, the Liouville-Arnold (LA) theorem  says that if a Hamiltonian system with
$2 n$ phase-space variables (positions and momenta), possesses $n$ constants of motion (including the
Hamiltonian)
which mutually commute among themselves, then the system is integrable and it possesses action-angle variables.

\hfill \break

\begin{definition}[label=def:A]
For  a more rigorous statement of the the theorem, along with its
proof, the reader is to referred to Chapter 11 of Ref.~\cite{fasano}.
\end{definition}

\hfill \break

\subsection{Application of the Liouville-Arnold theorem to the spinning BBH system}

To apply the LA theorem to the spinning BBH system, we need to determine
$2n$, the total number of positions and momenta. The total number of coordinate
appears to be 12; each vector $\vv{R}, \vv{P}, \vv{S}_1$ and $\vv{S}_2$
contributes 3 components. Despite that, $2n \neq 12$.
The reason we can't count all 12 components of vectors 
$\vv{R}, \vv{P}, \vv{S}_1$ and $\vv{S}_2$
is because when it comes to spins $\vv{S}_1$ and $\vv{S}_2$, 
it is not clear which components are positions and which are momenta. 
Remember, $2n$ is supposed to be the total number of positions and momenta.

To delineate the positions and momenta clearly in 
the vectors $\vv{R}, \vv{P}, \vv{S}_1$ and $\vv{S}_2$, 
we reproduce parts of
Eqs.~\eqref{C0-PBs_defined_1} and \eqref{C0-spin-PB} 
\begin{equation}
\left\{R_{i}, P_{j}\right\}=\delta_{ij} \quad \text { and }   \left\{\phi_{A}, S_{B}^{z}\right\}=\delta_{A B},    \label{canonical PB}
\end{equation}
which lets us see that $\phi_A$ (the azimuthal 
angle of $\vv{S}_A$) and $S^z_B$ (the $z$-component of 
$\vv{S}_B$) act like position and momentum, respectively.
 It is in this sense that we need to 
count positions and momenta to determine ``$2n$''
 for the application of the LA theorem.
Every spin vector thus contributes 2 positions-momenta.
Only 2 coordinates are needed to specify each spin
since spin magnitudes are constants: $\dot{S_A} = \pb{S_A, H} = 0$
(easily follows from Eqs.~\eqref{canonical PB} 
and \eqref{C0-PBs_defined_2}).
Therefore $2n$ for our spinning BBH
is $3 + 3+ 2+2 = 10$, which means that we require $10/2 = 5$ mutually commuting constants of motion
to establish integrability.

The PB between 
any two quantities
among $R^i, P_i, \phi^i_A, S^z_{iA}$ that falls outside the purview of Eqs.~\eqref{canonical PB} and \eqref{PBs_defined_2} is 0.

\newpage

\hfill \break

\begin{definition}[label=def:B]
Rigorously speaking,
integrability, action-angle variables and the LA theorem are built
on the foundations of symplectic geometry (a branch of differential geometry).
If the above process of counting the number of positions and momenta
for the  application of the LA theorem
seems shaky to the reader due to lack of rigor,
they are referred to 
\cite{jose, arnold, marsden_1, marsden_2} for an introduction to symplectic manifolds
and Darboux coordinates and the statement of the LA theorem against this mathematical backdrop.
\end{definition}

\hfill \break

For the spinning BBH, the five required 
commuting constants have been long-known \cite{Damour:2001tu}.
They are 
\begin{align}
H,~~~~~~~ J^2,~~~~~~~ L^2,~~~~~~~ J_z,~~~~~~~ \SeffL.    \label{CCs}
\end{align}
These
quantities have already been defined in Sec.~\ref{statement}.
Hence the 1.5PN spinning BBH is integrable and it possesses action-angle variables.

\chapter{Hamiltonian flow}

\section{Basics}

In this section, we will define curves and vector field following the 
approach adopted in Secs.~2.5, 2.6 and 2.7 of Ref.~\cite{schutz1980geometrical}.

\subsection{Curve}

\begin{figure}
   \centering
  \includegraphics[width=0.4\linewidth]{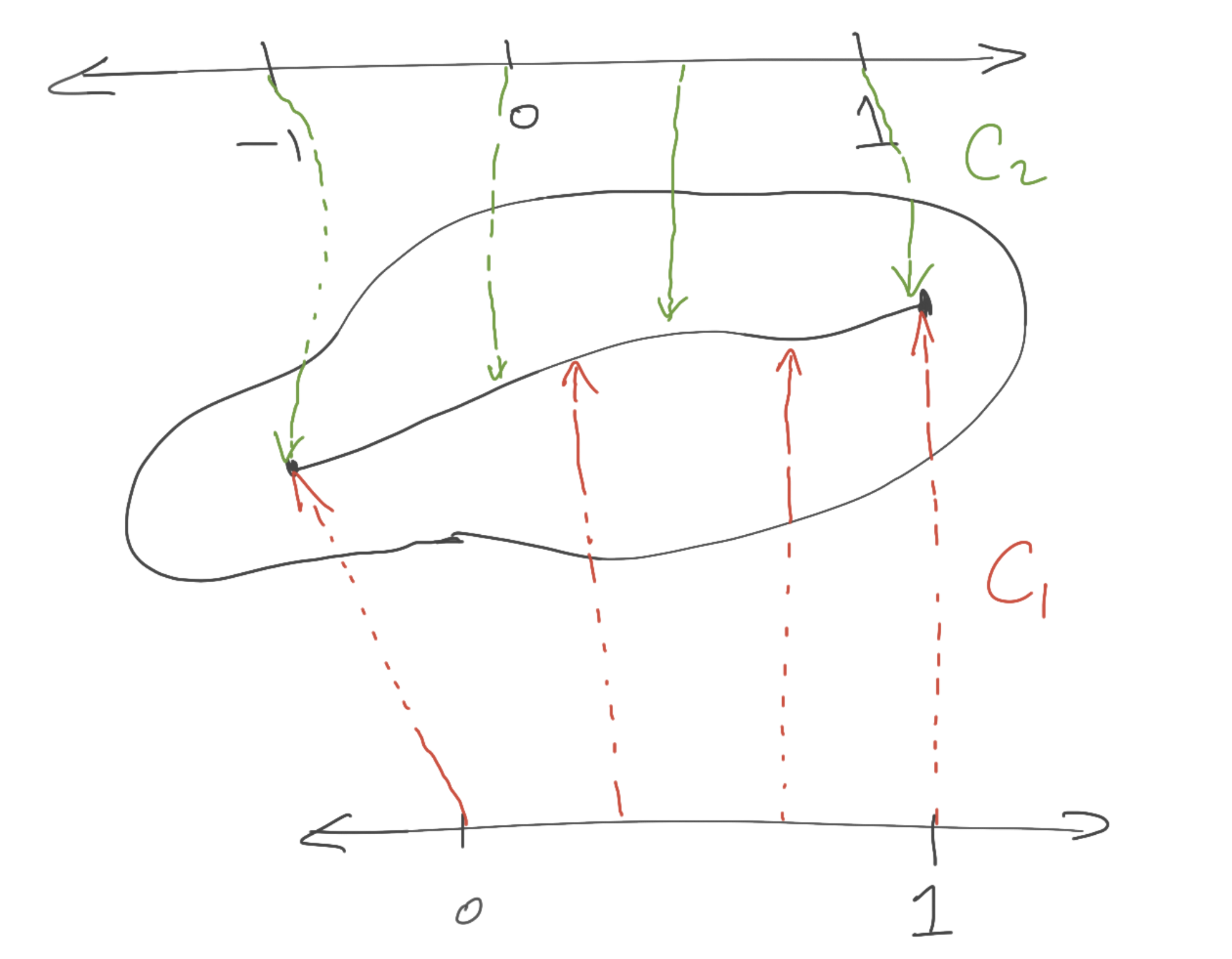}
  \caption{Pictorial  depiction of the concept of  a curve. 
  We are showing two different curves $C_1$ and $C_2$ which have the same image
  in manifold $M$.
    \vspace{-1.em}
  }
  \label{curve}
\end{figure}

A curve $C$ is a mapping from $\mathbb{R}$ to a manifold $M$, $C: \mathbb{R} \rightarrow  M$.
This is illustrated in Fig.~\ref{curve}. Note that two curves need not be the same 
even if they map to the same image in $M$. For example, the curves $C_1$ and $C_2$
are not the same in Fig.~\ref{curve}. This is so because their
pre-images (elements of the domain $\mathbb{R}$
corresponding to an image in $M$) are different.

\subsection{Vector field}

\begin{figure}
   \centering
  \includegraphics[width=0.4\linewidth]{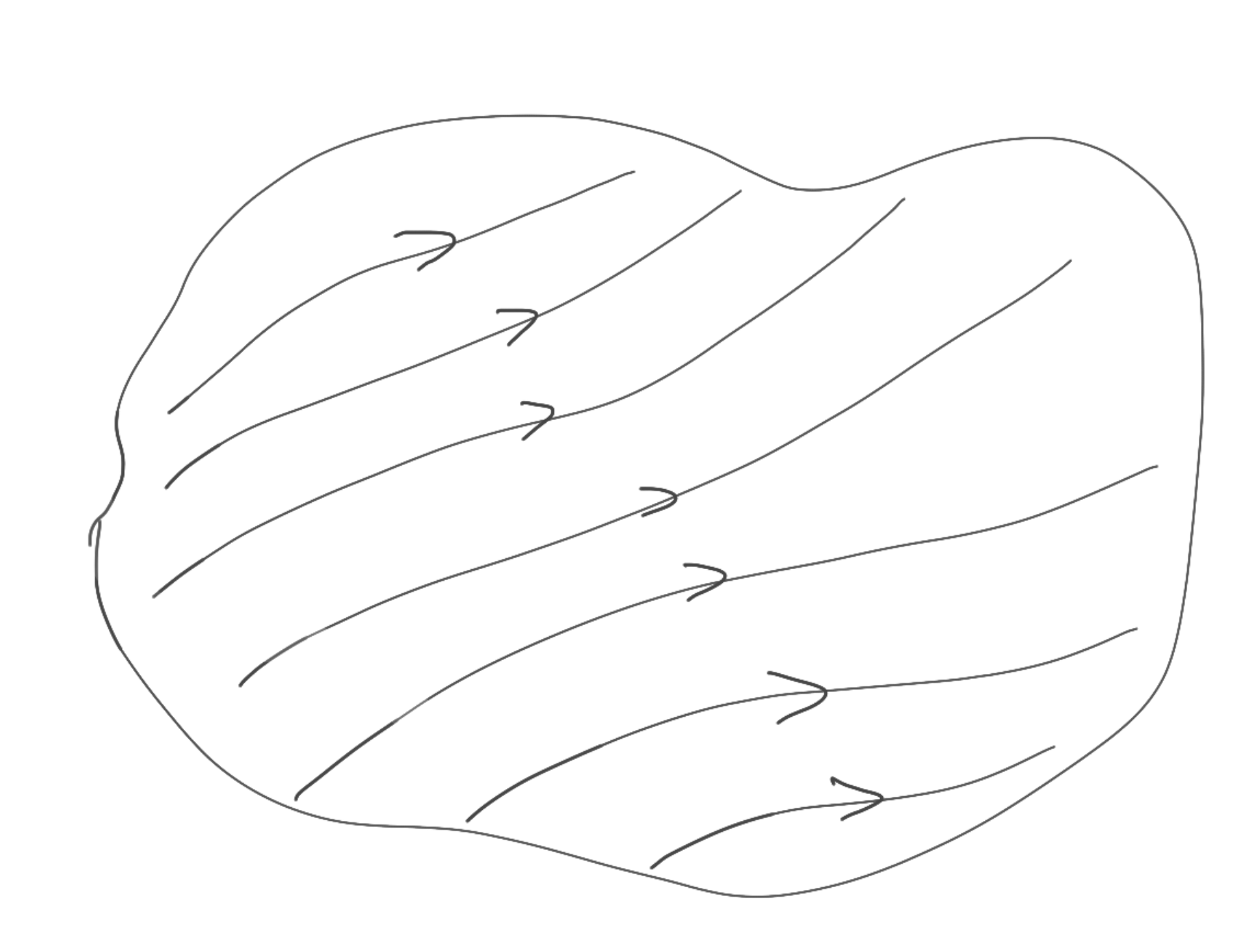}
  \caption{A manifold permeated with curves is a manifold with a vector field.
  These set of curves are also collectively refereed to as the flow of the
  associated vector field.
    \vspace{-1.em}
  }
  \label{vector_field}
\end{figure}

Imagine a curve $C$ defined by equations 
$\vv{x} = \vv{x}(\lambda)$ ($\vv{x}$ being the vector
of coordinates on $M$ and $\lambda  \in  \mathbb{R} $
 being the pre-image of $C$). Then at a point $P$ through which 
 the curve passes, we can 
define a vector whose components are $(dx^i/d \lambda)$. 
This vector (tangent to curve $C = x^i(\lambda)$)
is also denoted by $d/d \lambda$. After all,
 a vector is basically  a derivative operator.
This way we define vectors at every point of a curve $C$. If there are curves permeating
the manifold $M$, this can help us define a unique vector field on $M$, provided 
these curves don't intersect. A pictorial representation is shown in 
Fig.~\ref{vector_field}.

\section{Hamiltonian vector field and flow}

Unless stated otherwise, $\vv{V}$ from now on represents a column vector 
which contains all the components of $\vv{R}, \vv{P}, \vv{S}_1$ and $\vv{S}_2$.
Now, at all the points of the phase-space (which is a legitimate manifold),
we can define curves for a function $f(\vec{V})$ via (parameterized by $\lambda$)
\begin{align}
\frac{d \vv{V}}{d \lambda}   =  \pb{\vv{V},  f}  .       \label{H-flow}
\end{align}
It's implied that the above equation is to
 be interpreted in a component-wise manner.
The resulting vector field from these curves is called the 
\textit{Hamiltonian vector field} of $f$. Note that if $f = H$, then we will
have the Hamiltonian vector field of the 
Hamiltonian. More generally, with this definition, 
the Hamiltonian vector field of position 
$R_x$, momentum $P_z$ and the Hamiltonian $H$
are all well defined mathematical entities. 
For brevity we may refer to 
Hamiltonian flow as only the flow.

The \textsc{Mathematica} notebook for evaluating the Poisson 
brackets is available at (...Git) and (... YouTube).

\hfill \break

\begin{definition}[label=def:BB]
\textbf{Hamiltonian vector field:}
See Refs.~\cite{jose, arnold} for a more highbrow but
equivalent definition of Hamiltonian vector fields
using symplectic forms.
\end{definition}

\hfill \break

\begin{figure}
  \centering
  \includegraphics[width=0.4\linewidth]{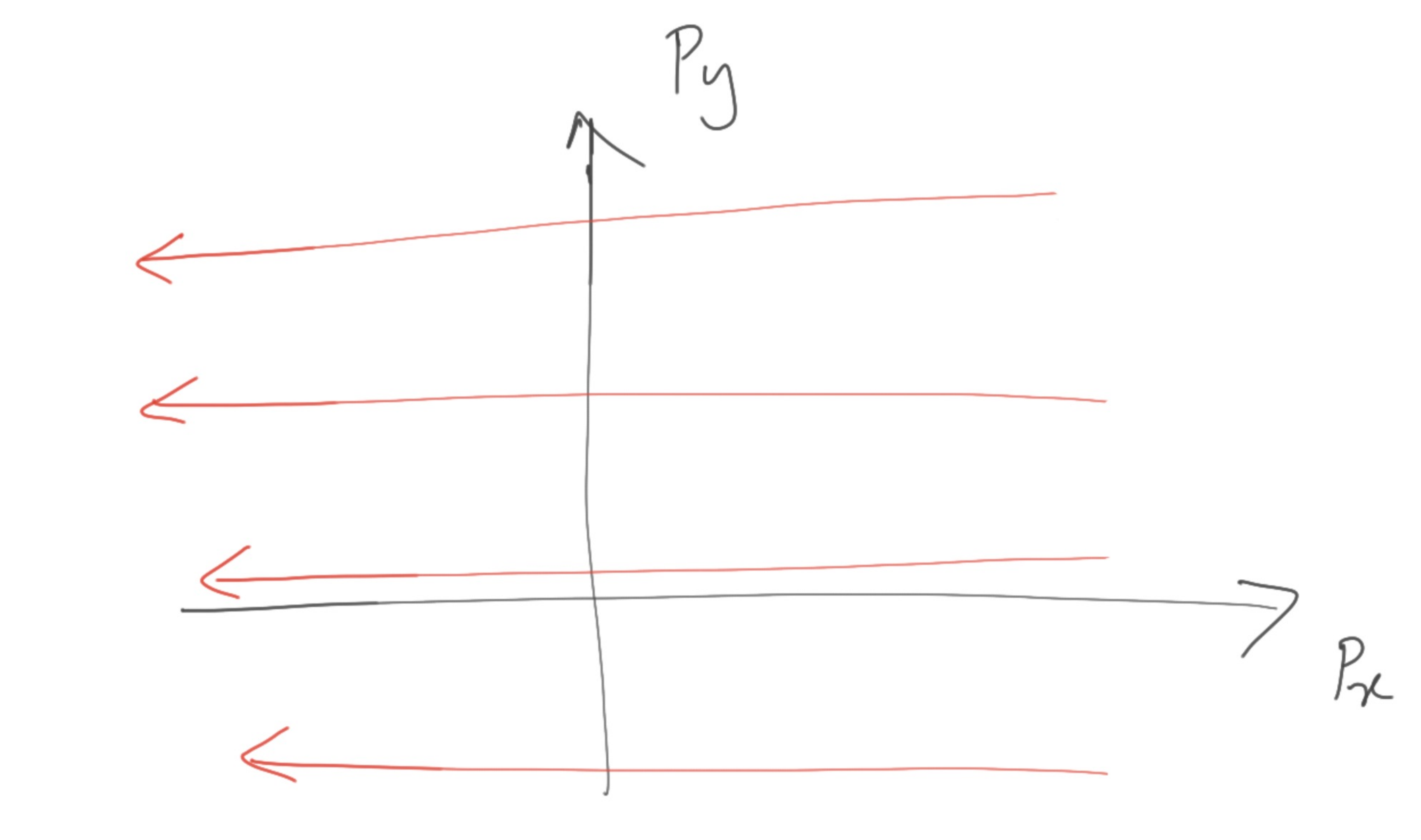}
  \caption{Pictorial representation of $R_x$ flow.
    \vspace{-1.em}
  }
  \label{Rx_flow}
\end{figure}

\begin{figure}
    \centering
  \includegraphics[width=0.4\linewidth]{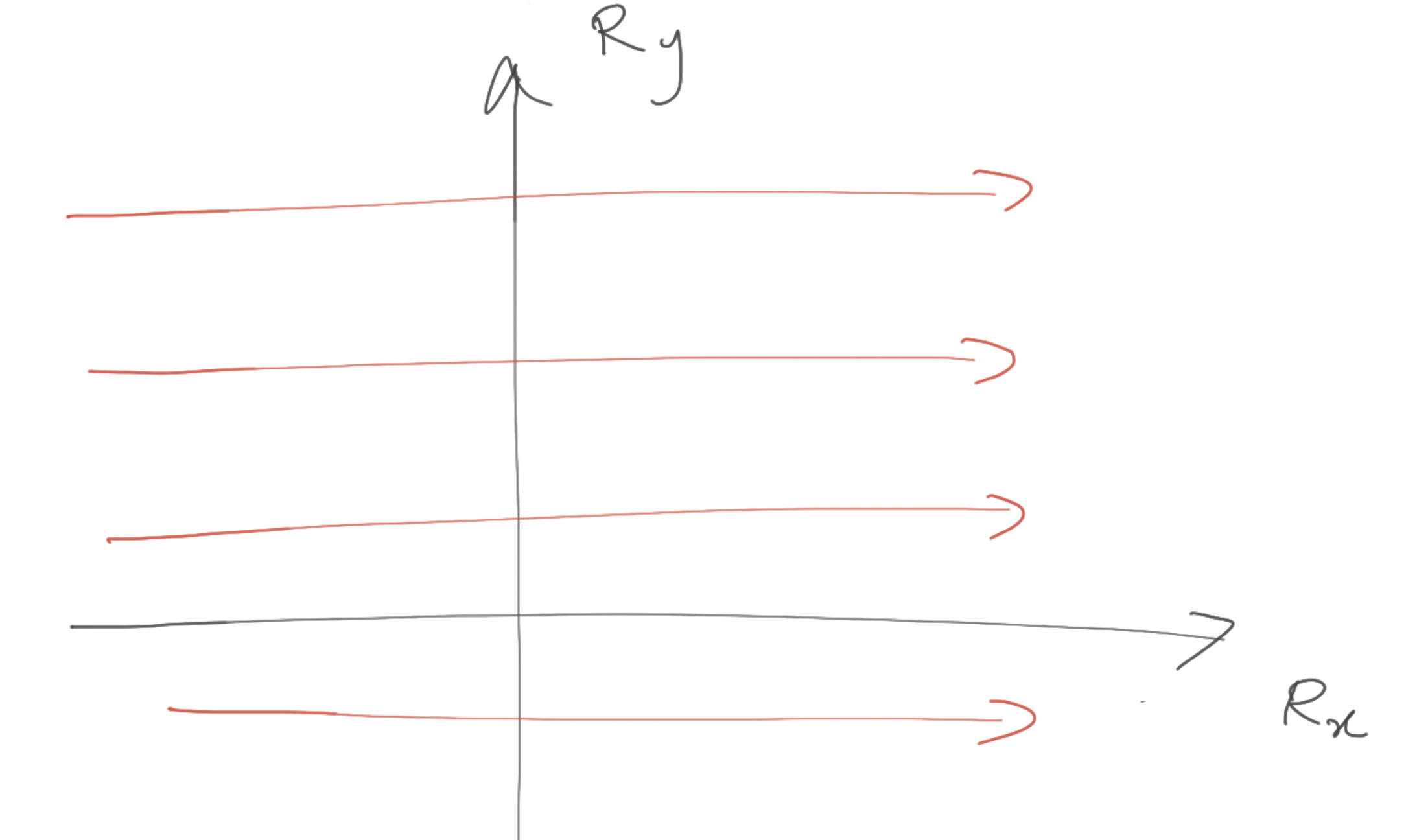}
  \caption{Pictorial representation of $P_x$ flow.
    \vspace{-1.em}
  }
  \label{Px_flow}
\end{figure}

\begin{figure}
    \centering
  \includegraphics[width=0.4\linewidth]{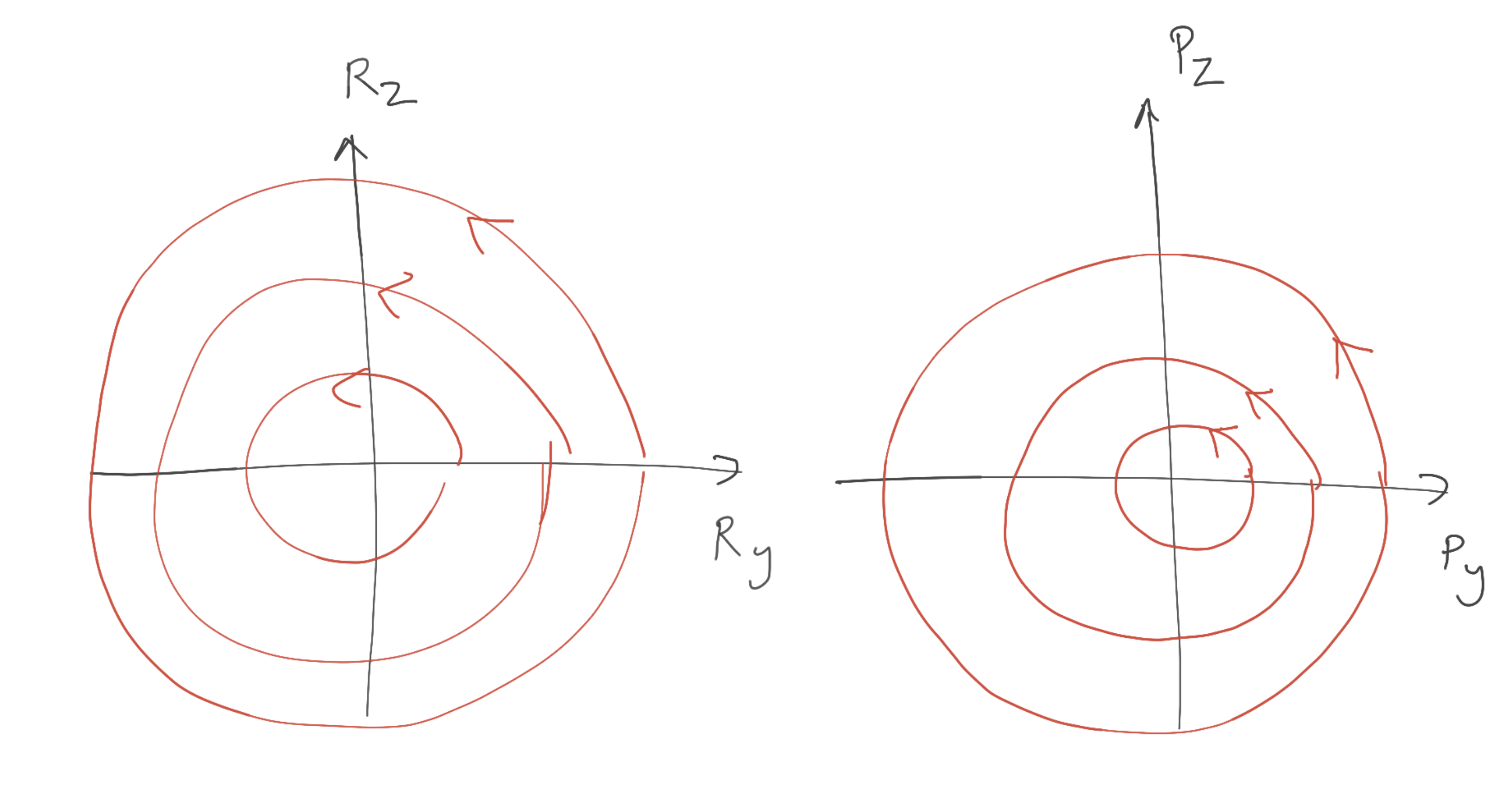}
  \caption{Pictorial representation of $L_x$ flow.
    \vspace{-1.em}
  }
  \label{Lx_flow}
\end{figure}

\begin{figure}
   \centering
  \includegraphics[width=0.4\linewidth]{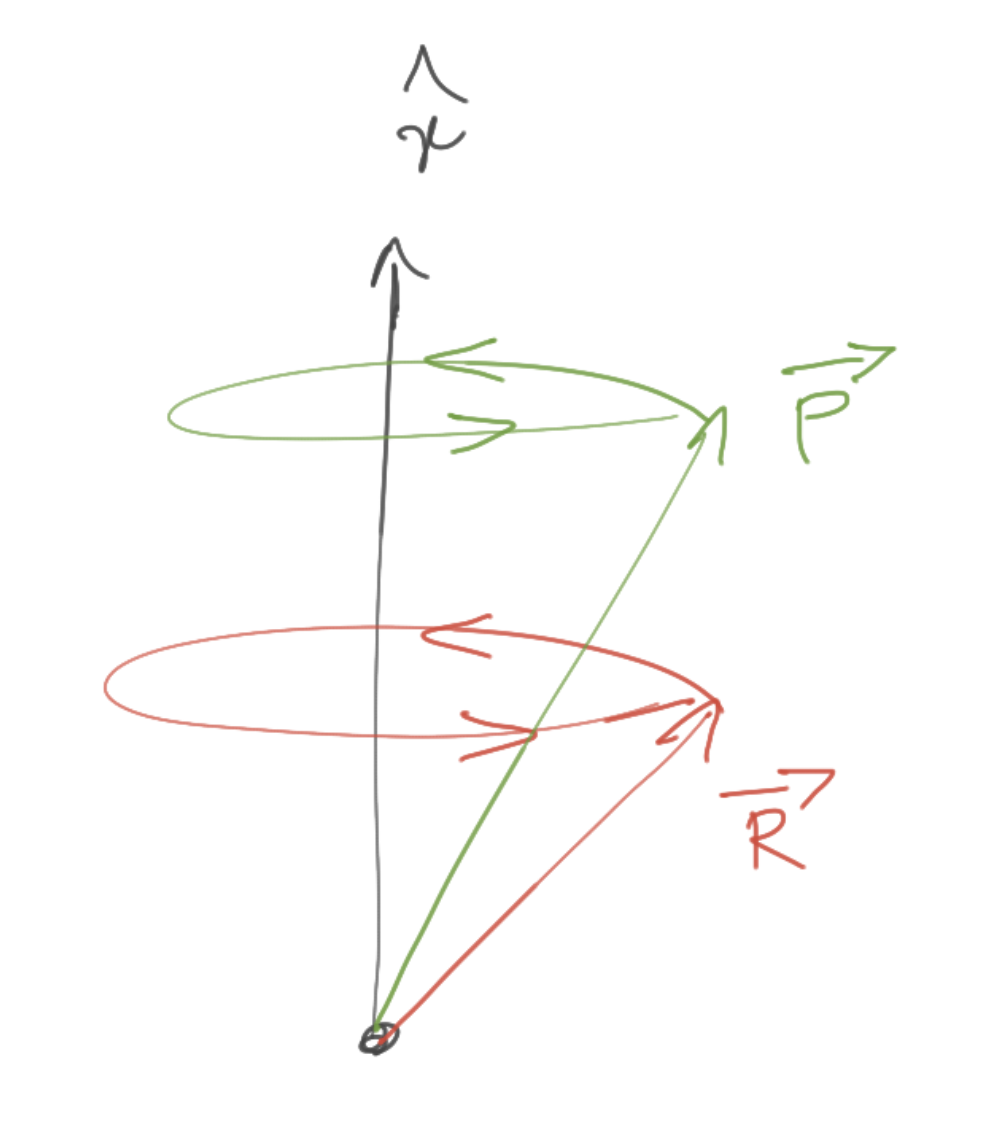}
  \caption{Pictorial representation of $L_x$ flow.
    \vspace{-1.em}
  }
  \label{Lx_flow_2}
\end{figure}

\begin{figure}
    \centering
  \includegraphics[width=0.4 \linewidth]{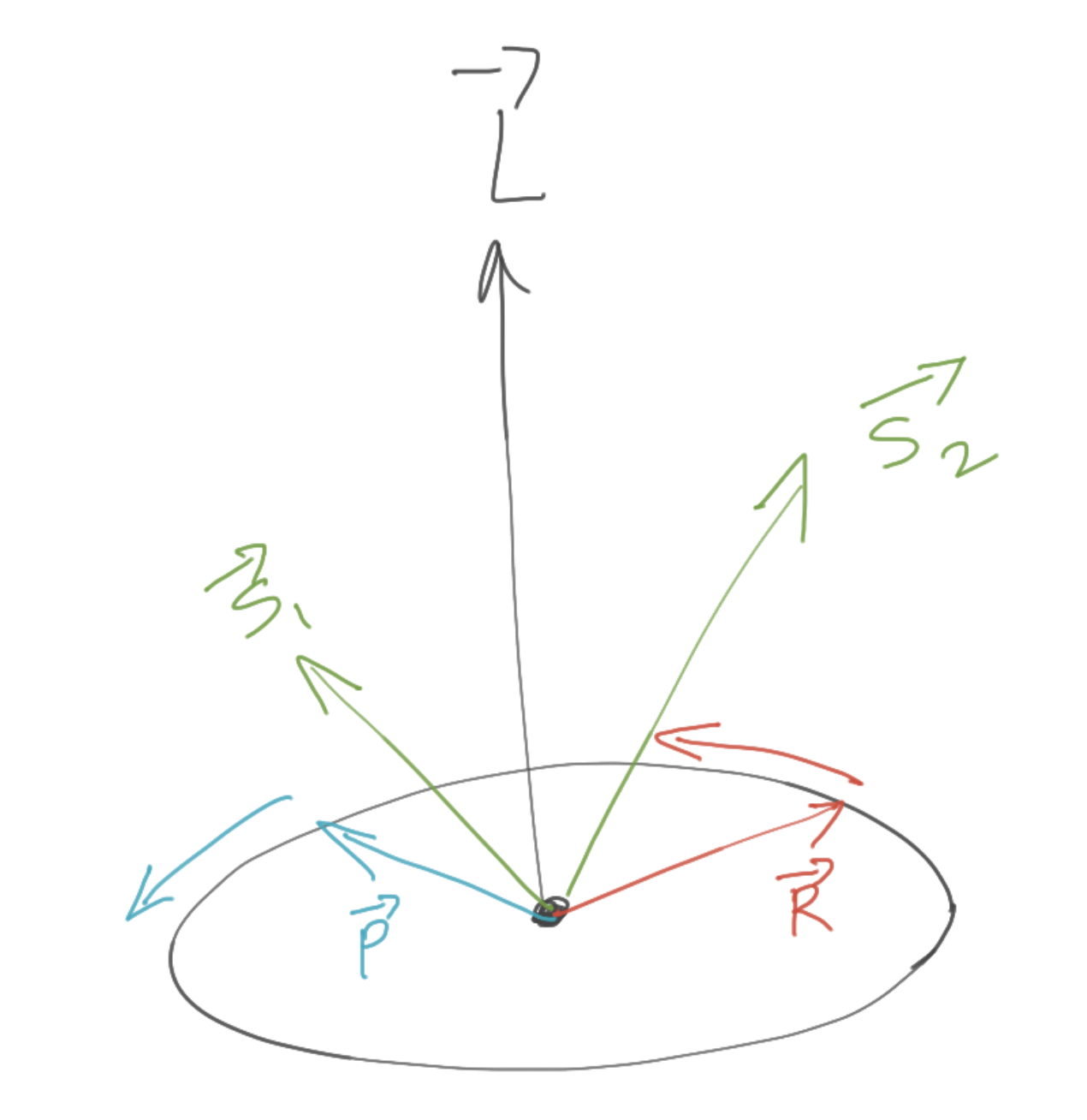}
  \caption{Pictorial representation of $L^2$ flow.
    \vspace{-1.em}
  }
  \label{Lsq_flow}
\end{figure}

\begin{figure}
  \centering
  \includegraphics[width=0.4\linewidth]{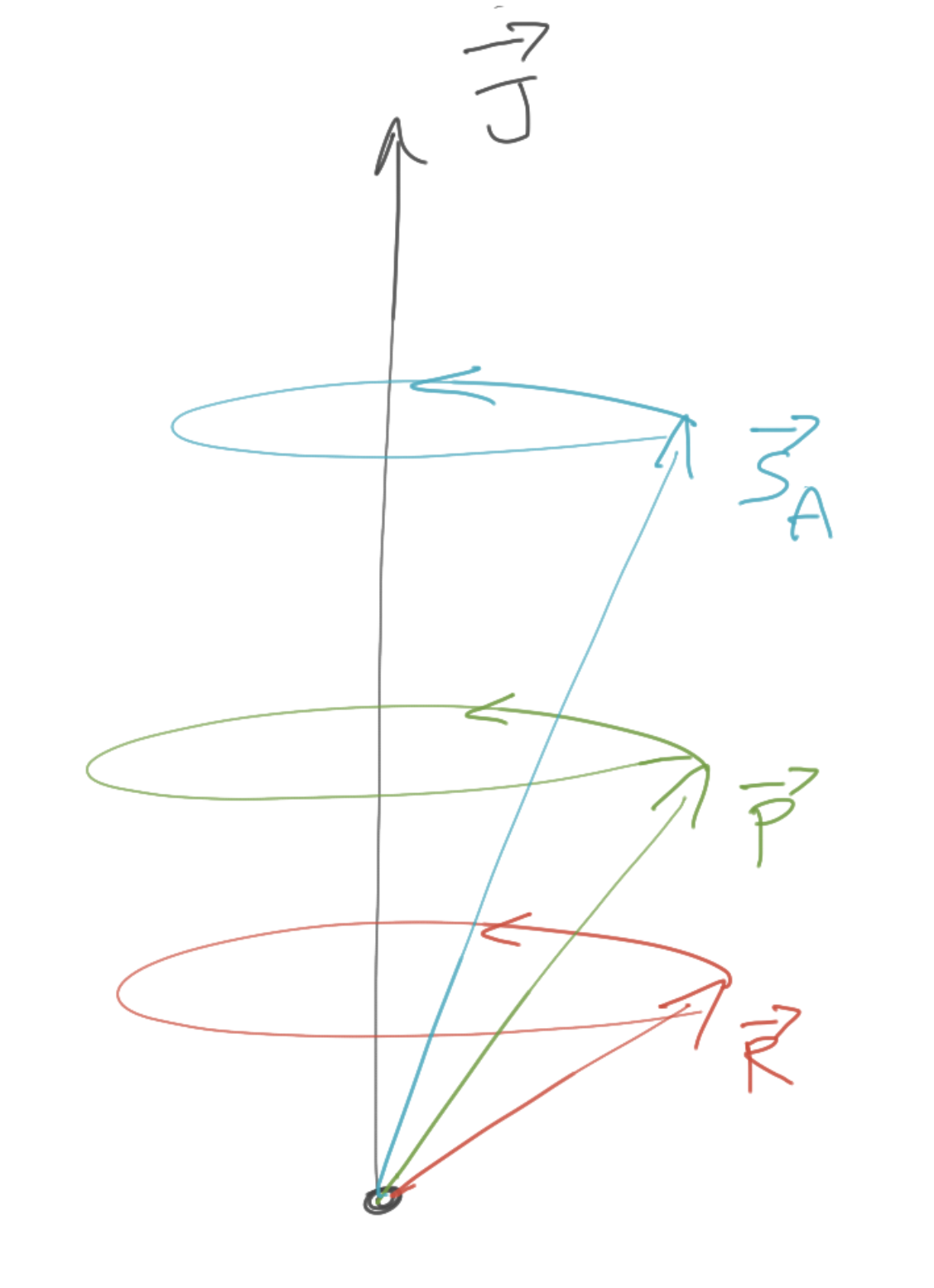}
  \caption{Pictorial representation of $J^2$ flow.
    \vspace{-1.em}
  }
  \label{Jsq_flow}
\end{figure}

\begin{Exercise}    \label{exercise-2}
\textbf{Problem:}  Draw pictures representing the Hamiltonian
flows of $R_x, P_x, L_x, L^2,$ and $J^2$.

\textbf{Solution:} 

For this exercise let $\vv{V}$ represent the totality of 
coordinates contained in $\vv{R}, \vv{P}, \vv{S}_1$ and $\vv{S}_2$,
unless stated otherwise.
We will give the mathematical
 solution for only the first flow (in addition to the
pictorial representation of the flow).
Solutions for the other flows will not be given; they are quite
simple to obtain.
Also the associated figures representing the flows,
the arrows denote the direction of
increasing flow parameter, i.e. $\lambda$.

\textbf{(a)}

Under the flow of $R_x$, among all the
coordinates of $\vv{V}$, only $P_x$ changes via
\begin{align}
\frac{d P_x}{d \lambda}  =  \pb{P_x, R_x} = -1,
\end{align}
since the PB of $R_x$ with all other variables is 0. 
The solution is $ P_x - P_x(\lambda_0) = (\lambda_0 -  \lambda).$
The  corresponding flow diagram is shown in
Fig.~\ref{Rx_flow}.\\

\textbf{(b)}

Under the flow of $P_x$, among all the
coordinates of $\vv{V}$, only $R_x$ changes via
\begin{align}
\frac{d R_x}{d \lambda}  =  \pb{R_x, P_x} = 1,
\end{align}
since the PB of $R_x$ with all other variables is 0. The 
corresponding flow diagram is shown in
Fig.~\ref{Px_flow}.

\textbf{(c)}

The flow of $L_x$ is encoded in 
\begin{align}
\frac{d \vec{R}}{d \lambda}=\hat{x} \times \vec{R},   \\
\frac{d \vec{P}}{d \lambda}=\hat{x} \times \vec{R},     \\
\frac{d \vec{S}_A}{d \lambda}=  0,   
\end{align}
that is to say that $\vec{R}$ and $\vec{P}$ rotate around $\hat{x}$,
whereas the spins don't move.
The  corresponding flow diagrams are shown in
Figs.~\ref{Lx_flow} and \ref{Lx_flow_2}. Note that the latter figure
is easier to understand because of the use of 3D vectors in the
drawing.

\textbf{(d)}

The flow of $L^2$ is given by
\begin{align}
\frac{d \vec{V}}{d \lambda}=\left\{\vec{V}, L^{2}\right\}=2 \vec{L} \times \vec{V}  \\ 
\quad \frac{d \vec{S}_{A}}{d \lambda_{1}}=0,
\end{align}
where in the above two equations, $\vv{V}$ stands for only
$\vv{R}$ and $\vv{P}$, and not the spin vectors.
This means that $\vv{R}$ and $\vv{P}$
rotate around $\vv{L}$ (which itself remains fixed since its PB
with $L^2$ is 0).
The  corresponding flow diagrams are shown in
Figs.~\ref{Lsq_flow}, where use of 3D vectors has been made, just like the 
previous case.

\textbf{(e)}

The flow of $J^2$ is given by
\begin{equation}
\frac{d \vec{V}}{d \lambda}=2 \vec{J} \times \vec{V}
\end{equation}
that is to say that all the four 3D vectors
rotate around $\vv{J}$ (which itself remains fixed since its PB
with $J^2$ is 0).
The corresponding flow diagrams are shown in
Figs.~\ref{Jsq_flow}.

\textbf{General remark:} Note that the flow diagrams show the trajectory 
of the flow but don't give us the information regarding 
the rate of flow. The rate has to be gleaned from the 
corresponding PB equations which encode the flow.

\end{Exercise}

\section{Hamiltonian flow of the Hamiltonian}

Hamiltonian flow of the Hamiltonian is encoded in 
\begin{align}
\frac{d \vv{V}}{d \lambda}  =   \pb{\vv{V}, H},
\end{align}
which upon comparison with Eq.~\eqref{C0-EOM-PB} is found to
be the EOM of the system. The flow parameter $\lambda$ plays the role of
time $t$. In this sense, the Hamiltonian flow of the 
Hamiltonian is indeed special for the flow of the Hamiltonian
dictates the real time evolution of a system.

\chapter{Action-angle variables in more detail}

\section{Poisson bracket of canonical coordinates}

\textbf{Definition:} Canonical coordinates are the ones which obey Hamilton's equations. \\
\textbf{Theorem:}  Any general canonical coordinates $(\vec{r}, \vec{p})$
 have the following PBs.
\begin{align}
\left\{p_{i}, p_{j}\right\}=\left\{r_{i}, r_{j}\right\}=0, 
&&\left\{r_{i}, p_{j}\right\}=\delta_{ij}.      \label{canonical_coordinates}
\end{align}
Also, any set of coordinates obeying the above PBs are canonical coordinates.

\hfill \break

\begin{definition}[label=def:CC]
See Theorem 10.17 of Ref.~\cite{fasano} for a proof of the above statement.
\end{definition}

\hfill \break

\section{Constructing action-angle variables}   \label{construct_AA}

In this section we will try to form a strategy to construct
 the action-angle coordinates
which satisfy the definition given in Sec.~\ref{define_integrable_sys}.
The definition of action-angle variables necessitates that the 
action-angle variables satisfy Eqs.~\ref{canonical_coordinates}.
We will break down our process of forming this strategy 
of constructing action-angle variables into a few steps. \\

\textbf{Step 1:}

\textbf{Theorem:} Consider the following integral
\begin{align}
\mc{J}_i  =   \frac{1}{2 \pi} \oint_{\gamma_i}   \vv{P} \cdot  d\vv{Q},       \label{action_integral}
\end{align}
where the line integral is done over 
a loop $\gamma_i$ which is on the 
$n$-dimensional sub-manifold
defined by the constant values of the $n$ commuting constants.
The flow under $\mc{J}$ by an amount $2 \pi$ forms another closed loop
in the phase space on the  above mentioned $n$-dimensional sub-manifold.

\begin{figure}
  \centering
  \includegraphics[width=0.7\linewidth]{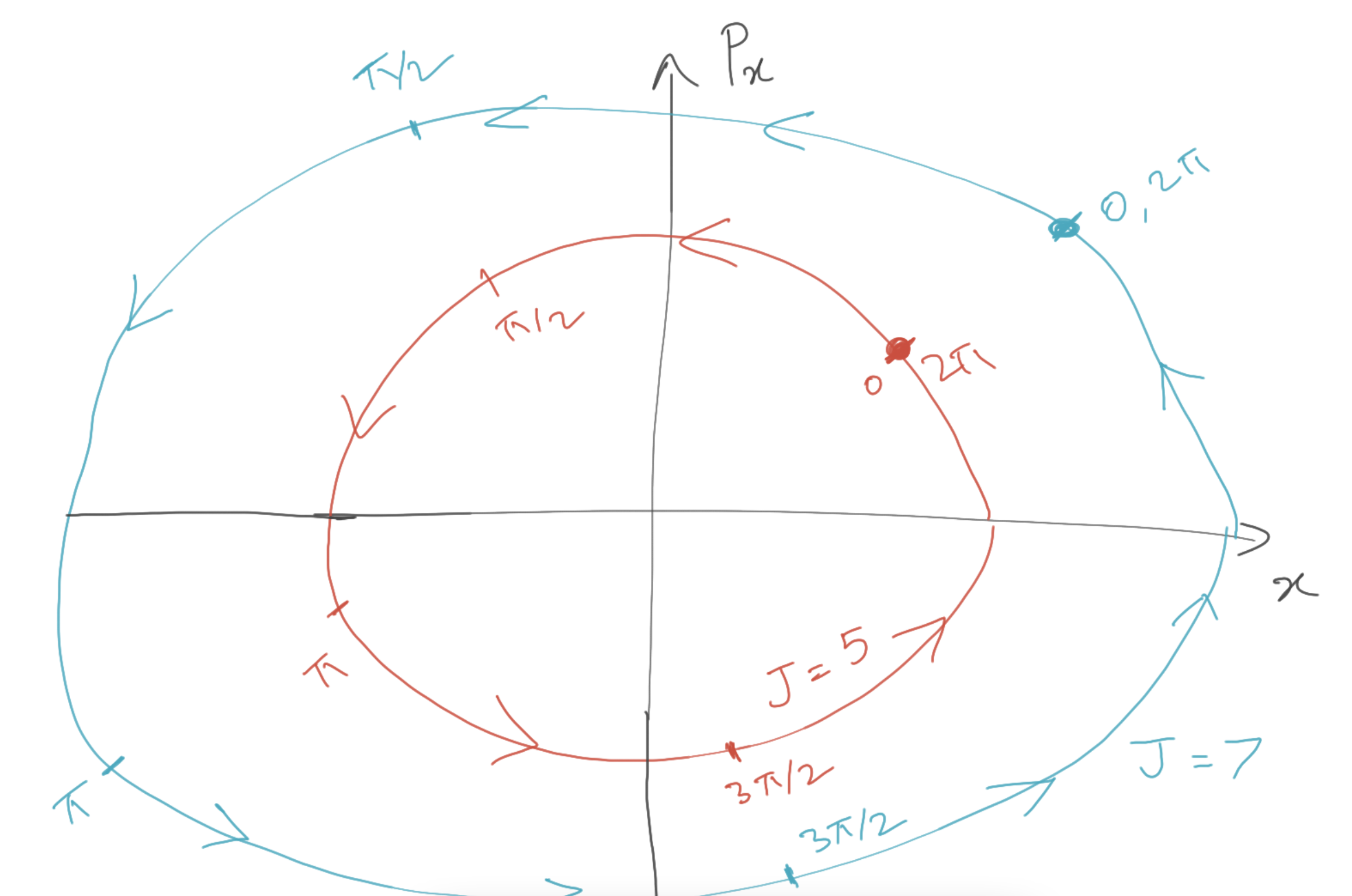}
  \caption{ Action flow makes a loop. On this loop,
  stamp the angle coordinates via $d \theta_i = d \lambda_i$,
  where $\lambda_i$ corresponds to the flow parameter under the 
  $\mc{J}_i$.
    \vspace{-1.em}
  }
  \label{action_loop}
\end{figure}

See Definition 11.6 and Theorem 11.6 of
Ref.~\cite{fasano} for a proof of this statement.
This is pictorially shown in
Fig.~\ref{action_loop}. 
We will argue later that this integral
is the action variable.
Many textbooks take this expression of action as a definition \cite{goldstein2013classical}. \\

\hfill \break

\begin{definition}[label=def:C2]

Note that there are two different loops in the picture in the  
above discussions. The first is the $\gamma_i$ loop (Loop 1) over which
the action integral is evaluated, and the other is the one which is generated
by flowing under the action (Loop 2).\\

In general, these two loops are not the same.
Just like Loop 1,
Loop 2 is also in the $n$-dimensional submanifold
defined by the constants values of $C_i$'s, since
an action flow does not change the values of the $C_i$'s.
 Also, it can be shown that the action
integral evaluated on both the loops is the same.

\end{definition}

\hfill \break

\newpage

\textbf{Step 2:} 

\textbf{Theorem:} The $\mc{J}_i$ constructed above are mutually commuting, 
\begin{equation}
\left\{J_{i}, J_{k}\right\}  =  0.          \label{actions_commute}
\end{equation}

\textbf{Proof:}  
Let's denote by $\vec{C}$, the vector of all $n$ commuting constants.
Now all the $\mc{J}_i$'s can be considered to be functions of only the $C_i$'s,
(apart from some other constants like the masses of the BHs).
See the proof of Proposition 11.2 of Ref.~\cite{fasano} for a proof
of this statement.

If that is so then it follows that
\begin{equation}
\left\{J_{i}, J_{k}\right\}=\sum_{l, m=1}^{n} \frac{\partial J_{i}}{\partial C_{l}} \frac{\partial J_{k}}{\partial C_{m}}\left\{C_{l}, C_{m}\right\}=0 . 
\end{equation}
In the above proof, we have used the chain rule for PBs (introduced in
Eq.~\eqref{C0-PBs_defined_2}). So, we have succeeded in ensuring the first equality of Eqs.~\eqref{canonical_coordinates}
to hold true. (with $\vv{p} = \vv{\mc{J}}$). We now move on to ensure the last equality of 
Eqs.~\eqref{canonical_coordinates} holds true. We are talking about i.e. $\left\{r_{i}, p_{j}\right\}=\delta_{ij} $,
or rather $\left\{\theta_{i}, \mc{J}_{j}\right\} = \delta_{ij}$.   \\

\textbf{Step 3:} Construct the angle coordinates $\theta_i$'s the following way.
Dictate that the way to increase the angle $\theta_i$ (associated to $\mc{J}_i$
via $\pb{\theta_i, \mc{J}_j} = \delta_{ij}$),
and keeping other (action-angle) coordinates fixed
is to flow under $\mc{J}_i$, thus tracing a loop.
Also, demand that under this flow, $d \theta_i = d \lambda$.

Now, is this construction consistent with the action-angle 
PB $\pb{\theta_i, \mc{J}_j} = \delta_{ij}$?
Yes. To see this, let $f = \mc{J}_i$ and $\vv{V}$ be $\theta_i$
 in Eq.~\eqref{C2-H-flow}, which then becomes
\begin{align}
\frac{d \theta_i  }{d \lambda}   = \pb{  \theta_i ,  \mc{J}_i}  = 1  .  
\end{align}
Note that we equated the two quantities to $1$
because of our demand $d \theta_i = d \lambda$.
Under the same flow of $\mc{J}_i$, we also have 
\begin{align}
\frac{d V  }{d \lambda}   = 0  ,      
\end{align}
where $V$ stands for any of the action variables or any of the
angle variables (except for $\theta_i$). This is
because of Eq.~\eqref{actions_commute} and also the fact that
we dictated that
one and only one angle $\theta^i$ will change as we flow under $\mc{J}_i$.
Note that this way of construction of angle coordinates 
applies only on the $n$-dimensional submanifold defined by the constant
values of the commuting constants. We have not specified 
how to construct angle coordinates off this submanifold.
Thus, the last equality of Eq.~\eqref{canonical_coordinates} is ensured 
by our construction of angle coordinates.\\

\textbf{Step 4:} We won't try to ensure the second equality of 
Eq.~\eqref{canonical_coordinates}, i.e. 
$\pb{ \theta_i, \theta_j} = 0$, because flow 
under any of the $\theta_i$'s
implies changing the corresponding action 
($\theta_i,  \mc{J}_j = \delta_{ij}$),
and for real-time evolution (flow under the Hamiltonian), actions 
do not change (see Eq.~\eqref{C1-AA_eqn_1}). So there is not much use
in stamping the angle coordinates off the 
$n$-dimensional submanifold defined by the constant values of the 
actions or the commuting constants.\\

\textbf{Step 5:} One might worry that we can have infinite number of 
$\mc{J}_i$' since we can have infinite number of loops $\gamma_i$'s
over which the integral in Eq.~\eqref{action_integral} is to be
performed. This is in conflict with our expectation that the number of 
actions and angles both has to be $n$, so that $n+n = 2n$ is the
dimensionality of the phase-space.

Actually, this is not a cause of concern. No matter how many action
integrals we compute using however many loops $\gamma_i$'s, there
will be only $n$ independent action variables. The rest will be linear 
combinations of these $n$ independent action variables. 
See Proposition 11.2 and 11.3 of Ref.~\cite{fasano} for a proof of this.\\

\textbf{Step 6:}
Now let's try to see if the action-angle variables defined or constructed
above match with the original definition given in Sec.~\ref{C1-definition}
or not. We will again not work things out from scratch but rather
outline the steps while referring to other sources.

We have already mentioned that
all the $\mc{J}_i$'s can be considered to be functions of only the $C_i$'s,
(apart from some other constants like the masses of the BHs).
All the $C_i$'s can also be considered to be functions of the
$\mc{J}_i$'s, i.e. $\mc{J}_i(\vec{C})$ is an invertible function,
if the determinant of the Jacobian of transformation 
(via the inverse function theorem)
is non-zero
\begin{align}
\operatorname{det}\left(\frac{\partial J_{i}}{\partial C_{j}}\right) \neq 0 .  \label{invert_condition}
\end{align}
The determinant is indeed non-zero for usual 
configurations. Now all this 
implies that the Hamiltonian (one of the $C_i$'s) is a function of
only the actions (in usual circumstances). This is
one of the defining criteria of action variables
(as per the definition given in Sec.~\ref{C1-definition}).

\hfill \break

\begin{definition}[label=def:C3]

\begin{center}
\textbf{Inverse of a function}
\end{center}

There needs to be some clarification in the context of 
the meaning of the $\mc{J}_i$'s being functions of $C_j$'s, 
and vice-versa. To simplify matters, lets talk in terms of 
a function $f$ of  a single variable $g$.\\

Note that $f$ being  a function of $g$ in the neighborhood
of point $g=g_0$ does not imply that 
$f$ has to be expressible in terms of $g$ in closed-form
using standard functions like sine, cosine, exponential or elliptic functions.
$f$ being  a function of $g$ in the neighborhood
of point $g=g_0$ means is that the function
$g(f)$ is an injective function in this neighborhood, so that
the inverse $f(g)$ is clearly defined. It may take numerical
root-finding to evaluate $f(g)$ but that's alright. 
See articles on implicit function theorem and inverse 
function theorem for more details. \\

For example, with the famous Kepler equation $l = u - e \sin u$,
$l$ is a function of $u$ (clearly), but $u$ is also a function of $l$
for all $l$'s since $l(u)$ is an injective function. Also, 
to evaluate $u(l)$, numerical root-finding is required. \\

Yet another example is $y= \sin x$, which does not have an inverse function
in a neighborhood around $x= \pi/2$ because in this neighborhood, 
$y(x)$ is not injective due to the maxima at $x= \pi/2$. So, one can't 
define $x(y)$ in this neighborhood, although one can clearly define 
$x(y)$ in small neighborhood around some other point say $x = \pi/4$.\\

So, for a one variable function $y(x)$,
 the inverse function does not exist in a 
neighborhood containing  a point $x_0$, where $dy/dx(x=x_0) = 0$. The 
generalization of this to multi-variable case is that the determinant
of the corresponding Jacobian has to be non-zero for the multi-variable
functions to be invertible (the inverse function theorem).
We apply precisely this theorem in Eq.~\eqref{invert_condition}.

\end{definition}

\hfill \break

Now let's come to the other criterion of this definition of action-angle variables.
The other criterion demands that 
{$\{ \vec{p}, \vec{q} \}(\theta_i + 2 \pi)  = \{ \vec{p}, \vec{q} \}(\theta_i ) $},
i.e. $\vv{p}$ and $\vv{q}$ are $2 \pi$-periodic
functions of $\theta_i$'s. This is clearly obvious from our construction of angle 
variables. As per this construction, changing only one of the angles
is tantamount to flowing under the corresponding action and doing so by an
amount $2 \pi$ brings us back to where we started from, thus forming  a loop. 
Hence, {$\{ \vec{p}, \vec{q} \}(\theta_i + 2 \pi)  = \{ \vec{p}, \vec{q} \}(\theta_i ) $} is indeed satisfied.

All in all, the take-home message is that (also shown in Fig.~\ref{action_loop})\\
\begin{tcolorbox}
as we flow under $\mc{J}_i$ (one of the actions), we form a loop after flowing by $\Delta \lambda_i = 2 \pi$.
We can also stamp the angle coordinate $\theta^i$ on this loop by setting $\theta_i = \lambda_i + C_0$, where 
$C_0$ is some constant real offset.
\end{tcolorbox}

\section{Action-angle variables of a simple harmonic oscillator}

Without explicit calculations, we state that for a SHO, the action-angle
variables are (with $\omega_0 =  \sqrt{k/m}$ and $\theta_0 \in \mathbb{R}$)
\begin{align}
\mc{J}  &= \frac{H}{\omega_0}  ,   \\
\theta    & = \arctan \frac{m \omega_0 q}{p} + \theta_0  .
\end{align}
The corresponding figure is again Fig.~\ref{action_loop}. Larger action
values means larger energy $H$ and hence a larger loop in the phase-space,
as is evident from the figure.

\section{How to flow under the actions?}

\begin{figure}
  \centering
  \includegraphics[width=0.8\linewidth]{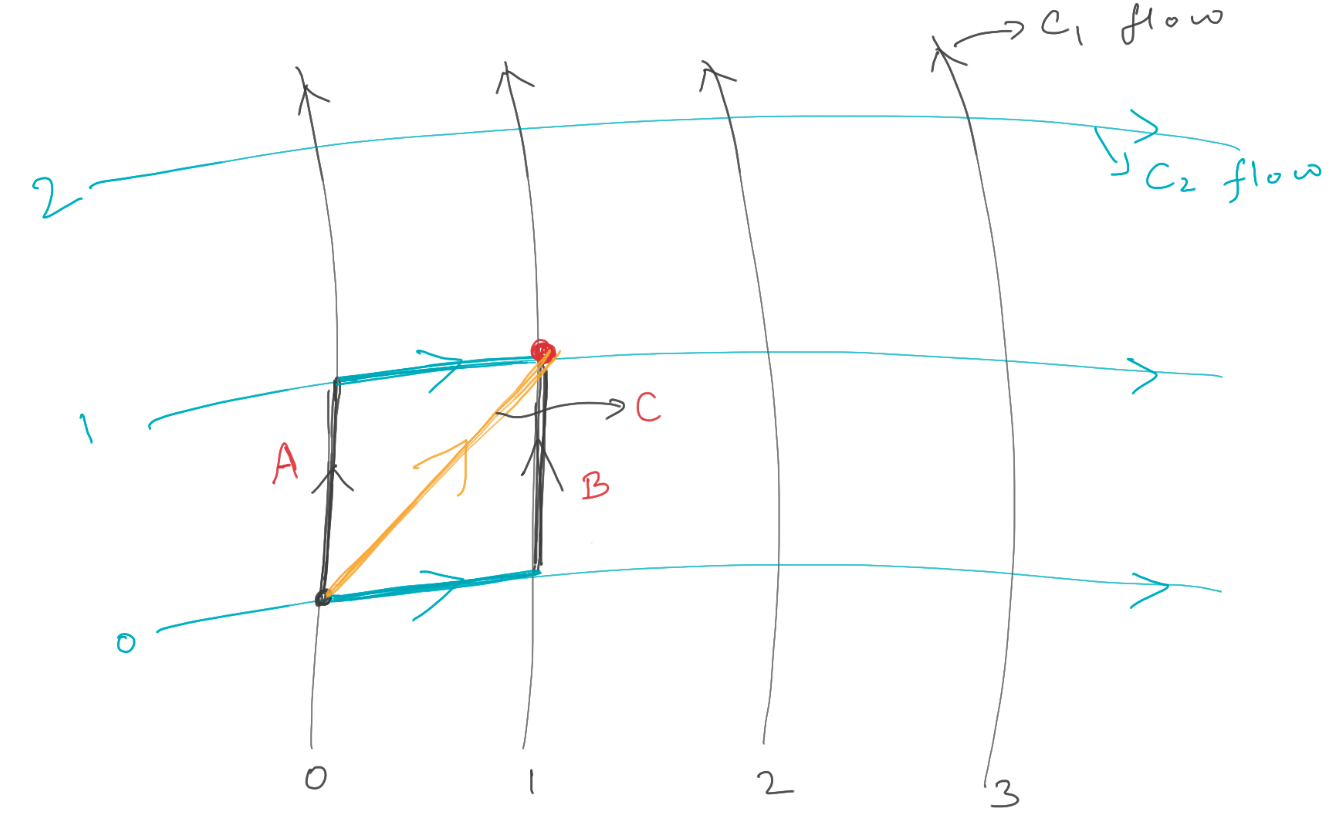}
  \caption{ Setting up coordinates on the manifold using
  the flows of the commuting constants.
    \vspace{-1.em}
  }
  \label{flow_coordinates}
\end{figure}

\subsection{Flowing under commuting constants}    \label{flow_breakdown}

From Secs.~2.15 and 2.16 of Ref.~\cite{schutz1980geometrical},
or Sec.~9.6 and Exercise 9.9 of Ref.~\cite{misner2017gravitation},
we learn that vector fields of commuting quantities can be used to
set up coordinates on the manifold,
the flow parameter being the coordinate. We have already seen 
the application of this above
when we set up the angle coordinates along the flow of the
corresponding actions in Sec.~\ref{construct_AA}. 
Fig.~\ref{flow_coordinates}
is a pictorial depiction of setting up coordinates this way.

Now we state two theorems:
\begin{itemize}
\item \textbf{Theorem:} Order of flow under two commuting quantities
(need not be constants) does not matter.

Pictorially, flowing under two commuting quantities $C_1$ and $C_2$ 
in Fig.~\ref{flow_coordinates}
by fixed amounts in different orders (paths $A$ and $B$) 
starting from the same point 
point (black dot) leads us to the same ending point (red dot).

If we take for granted that vector fields of commuting 
quantities can be used to
set up coordinates on the manifold, with the 
flow parameter being the coordinate (see the 
beginning of Sec.~\ref{flow_breakdown}),
then the proof of this theorem
becomes almost self-evident from Fig.~\ref{flow_coordinates}.

\item \textbf{Theorem:} If $\pb{C_1, C_2} = 0$, then a flow under 
$C_1 + C_2$ by a certain amount $\Delta \lambda$ is equivalent 
to a flow under one of them followed by the other, each by an amount
$\Delta \lambda$.

Here is the proof. Let $\vec{V}^i$ denote one of the 
components of $\vec{R}, \vec{P}, \vec{S}_1$ and $\vec{S}_2$. Let
$\lambda_1$ and $\lambda_2$ denote the coordinates generated $C_1$
and $C_2$ flows as shown in Fig.~\ref{flow_coordinates}. So, we have
$V^i = V^i(\lambda_1, \lambda_2)$ and from basic multivariable
calculus
\begin{align}
& dV^i    = \frac{\pd V^i}{\pd \lambda_1} d \lambda_1 + \frac{\pd V^i}{\pd \lambda_2}   d \lambda_2   ,\\
&  \frac{dV^i}{d \lambda}    = \frac{\pd V^i}{\pd \lambda_1} + \frac{\pd V^i}{\pd \lambda_2}     \quad \quad  \text{if~}d\lambda_1 = d \lambda_2 = d \lambda  ,  \\
& \frac{dV^i}{d \lambda}   =  \pb{V^i , C_1} +  \pb{V^i , C_2} = \pb{V^i , C_1 + C_2} ,
\\
\implies  &  \frac{\pd V^i}{\pd \lambda_1} d \lambda + \frac{\pd V^i}{\pd \lambda_2}   d \lambda   =    \pb{V^i , C_1 + C_2}  d \lambda  .    \label{simul_flow}
\end{align}
Now the LHS of the above equation denotes the change in going from the
black dot to the red dot in Fig.~\ref{flow_coordinates},
which can be accomplished by flowing under $C_1$ and $C_2$ in any order 
(as already discussed after stating the previous theorem above).
This, along with Eq.~\eqref{simul_flow} finally proves the current theorem.
\end{itemize}

So, what we have learned is that  \\
\begin{tcolorbox}
\begin{itemize}
\item The order of flows under commuting quantities does not matter.
\item A simultaneous flow under commuting quantities 
by $\Delta \lambda$ can be broken down into the individual flows under them (by an amount $\Delta \lambda$ each).
\end{itemize}
\end{tcolorbox}

\subsection{Breaking down the action flow}      \label{action_flow_breakdown}

\begin{tcolorbox}
Let us assume two things for the rest of the material in this chapter
\begin{itemize}
\item We have all the actions in terms of the 
commuting constants: $\vec{J}(\vec{C})$. 
We will compute $\vec{J}(\vec{C})$ in Chapters \ref{C4-chapter-5}
and \ref{C5-chapter-6}.
\item We have closed-form solution for the flows under all the commuting
constants $C_i$'s. We will find these solutions for some
of the commuting constants in Chapter \ref{C4-chapter-5}.
The solution of the flows under the Hamiltonian and $\Eff$ 
is given in Refs.~\cite{Cho:2019brd} and \cite{tanay2021action}, 
respectively.
\end{itemize}
\end{tcolorbox}

With the above two assumptions, how 
can we flow under an action?
The answer is the chain-rule property of the PBs 
(Eq.~\eqref{C0-PBs_defined_2}). So, we have
\begin{equation}     \label{action_flow_2}
\frac{d \vec{V}}{d \lambda}=\left\{\vec{V}, \mc{J}_{i}\right\}=\left\{\vec{V}, C_{j}\right\}\left(\frac{\partial \mc{J}_{i}}{\partial C_{j}}\right).
\end{equation}
If we have $\cJ(\vv{C})$, then the partial derivatives can be evaluated and
they are simply constants because they will be functions of the $C_i$'s.
Apart from that, if we have the solution for flow under all the commuting 
constants, then the solution of flow under any of the actions
can be easily had. Note that Eq.~\eqref{action_flow_2} is the
equation for simultaneous flow under multiple commuting constants.
But from Sec.~\ref{flow_breakdown}, we know that we can flow under 
all these commuting 
constants one-by-one (order doesn't matter) and get the solution
for the flow under $\mc{J}_i$ by any finite amount, provided 
the we have $\mc{J}(\vv{C})$ and flow solution under all the $C_i$'s.

\subsection{Computing frequencies}    \label{compute_freq}

How do we compute the frequencies $\omega_i = \pd H/\pd \mc{J}_i$ given 
$\mc{J}_i(\vec{C})$, the Hamiltonian $H$ being one of the $C_i$'s?
It's trivial to compute the Jacobian $\pd \mc{J}_i/\pd C_j$ using 
a computer algebra system as \textsc{Mathematica}.

Now, the frequencies $\omega_i = \pd H/\pd \mc{J}_i$ are elements of the 
the Jacobian of the inverse transformation, i.e. 
$\pd C_i/\pd \mc{J}_j$. From the inverse function theorem, we know 
that the Jacobian of the inverse transformation is the matrix 
inverse of the Jacobian of the original transformation.

\section{Constructing the action-angle based solution of the spinning BBH system}

Assuming that we have $\vec{\mc{J}}(\vv{C})$, and the solution for flow under
all the $C_i$'s, we now explicate the operational way to construct the
action-angle based solution of the system. $\vv{V}$ represents the totality of
the variables contained in the vectors $\vec{R}, \vec{P}, \vec{S}_1$ and $\vec{S}_2$,
and $\vv{V}_0$ denotes its initial value at time $t=0$.
We take $\vv{V}_0$ to correspond to all the angles being 0.

Suppose the solution is required at  a later time $t$. So, we know that
all the angles have changed to $\Delta \theta_i = \omega_i t$. 
We can easily have the numerical value of $\Delta \theta_i$ because we 
can compute all the $\omega_i$'s (see Sec.~\ref{compute_freq}).
Remember from Step 3 of Sec.~\ref{construct_AA}
that $\Delta \theta_i$, the change in the angles can be achieved by flowing 
under the corresponding actions by an equal amount
$\Delta \lambda_i = \Delta \theta_i$. The order of action flows does not matter
since $\pb{\mc{J}_i, \mc{J}_j} = 0$. Now the problem is reduced to flowing 
under all the $\mc{J}_i$'s by specified amounts, which we have
already figured out how to do in Sec.~\ref{action_flow_breakdown}.

\section{Afterthoughts and the plan ahead}

Now, the only missing ingredients towards constructing closed-form solutions
to the system are the $\vec{\mc{J}}(\vv{C})$ expressions and
the solutions of the flows under all the $C_i$'s.
We won't discuss the solutions of the flows under all the $C_i$'s, 
except for referring the reader to the relevant sources.
\begin{itemize}
\item The solution of the flow under $J^2, L^2$, and $J_z$ is very simple;
all the four 3D vectors 
$\vv{R}, \vv{P}, \vv{S}_1$ and $\vv{S}_2$ 
keep their magnitudes fixed and 
rotate around some fixed vector at a constant rate, with the exception that
spins don't move at all under the $L^2$ flow. We will discuss this 
to some degree in Chapter~\ref{C4-chapter-5}.
\item The solution of the flow under the Hamiltonian is worked out in 
Ref.~\cite{Cho:2019brd}. This paper ignores the 1PN Hamiltonian terms
for succinctness.
The solution is contained in Eqs.~(3.13, 3.24, 3.32, 3.42, 3.43)
of Ref.~\cite{Cho:2019brd}.

But we have already incorporated the 1PN terms of the Hamiltonian 
and implemented the flow solutions for $\vec{S}_1$ and $\vec{S}_2$ in a \textsc{Mathematica} package \texttt{BBHpnToolkit} \cite{MMA1}.
 The calculations behind this package will be presented in 
detail in an upcoming manuscript \cite{next_paper}.

\item The solution of flow under $\SeffL$ is worked out in 
Ref.~\cite{tanay2021action}.
The solution is contained in Eqs.~(A39, A65, A75)
of Ref.~\cite{tanay2021action}. Ref.~\cite{tanay2021action}
 does not explicitly give the solution
of $\vec{S}_1$ but it can be had in the 
same manner as that for $\vec{L}$ (Eq.~A65).
Then we can have $\vec{S}_2 = \vec{J} - \vec{L} -\vec{S}_1$.
\end{itemize}
So, in the remainder of these lecture notes, our focus will be on the only
remaining task which is to obtain $\vv{\cJ}(\vec{C})$.

\chapter{Computation of the first four actions}    \label{chapter-5}

Recall from Sec.~\ref{C3-construct_AA} that action variables are given by
\begin{equation}
\mathcal{J}_{i}=\frac{1}{2 \pi} \oint_{\gamma_{i}} \vec{P} \cdot d \vec{Q} , 
\end{equation}
where the integral is performed over  a loop on the $n$-dimensional 
submanifold defined by constant values of the $n$ commuting constants.
One way to remain on this submanifold is to flow under any of the 
$n$ commuting constants, for if you flow under any of the commuting 
constants, the other commuting constants don't change. This is because
under the flow of $C_j$, $C_i$ changes as (using Eq.~\eqref{H-flow})
\begin{align}
\frac{d C_i}{d \lambda} =  \pb{C_i, C_j}   = 0 .
\end{align}

Owing to this observation, we will try to
form loops for action integration while flowing 
under the commuting constants in the following sections.

\section{Computation of $\mc{J}_1$}    \label{result_first_action}

Let's flow under one of the commuting constants $J^2$, square of the 
magnitude of the total angular momentum $\vv{J} = \vv{L} + \vv{S}_1 + \vv{S}_2$.
Its flow equation is (with $\vv{V}$ standing for 
the column vector containing all the variables in $\vv{R}, \vv{P}, \vv{S}_1$
and $\vv{S}_2$)
\begin{equation}
\frac{d \vec{V}}{d \lambda} = 2 \vec{J} \times \vec{V},      \label{flow_action_1}
\end{equation}
which implies that all the four 3D vectors rotate around 
the fixed $2 \vv{J}$ vector (note that $\pb{\vv{J}, J^2} = 0$).
Using $\vv{n}$ to denote the vector around which all the four
3D vectors rotate, we see that $\vv{n} = 2 \vv{J} $.
In fact we had already worked this out and the pictorial 
representation has already been presented in the form 
of Fig.~\ref{Jsq_flow}.

Now, Eq.~\eqref{flow_action_1} implies that we will
arrive at where we started from (thus closing  a loop) after we have
flowed by an amount $\Delta \lambda = 
2 \pi/|\text{angular~velocity}|  
= 2 \pi/(2 J)   = \pi/J$.

Now, we break down the action integral as
\begin{equation}
\begin{aligned}
\mathcal{J} &=\mathcal{J}^{\text {orb }}+\mathcal{J}^{\text {spin }} \\
\mathcal{J}^{\text {orb }} & \equiv \frac{1}{2 \pi} \oint_{\mathcal{C}} \sum_{i} P_{i} d R^{i}   \\
\mathcal{J}_{A}^{\mathrm{spin}}  & =\frac{1}{2 \pi} \oint S_{A}^{z} d \phi_{A} .
\end{aligned}
\end{equation}
Let's tackle the orbital sector first. The orbital contribution
to the action integral becomes
\begin{align}
\mathcal{J}^{\text {orb }} &=\frac{1}{2 \pi} \int_{0}^{\Delta \lambda} P_{i} \frac{d R^{i}}{d \lambda} d \lambda=\frac{1}{2 \pi} \int_{0}^{\Delta \lambda} \vec{P} \cdot(\vec{n} \times \vec{R}) d \lambda \\
&=\frac{1}{2 \pi} \int_{0}^{\Delta \lambda} \vec{n} \cdot \vec{L} d \lambda=\hat{n} \cdot \vec{L}   .     \label{orbital_contribution}
\end{align}
The spin sector integral is 
\begin{align}
\mathcal{J}_{A}^{\mathrm{spin}}  & =\frac{1}{2 \pi} \oint S_{A}^{z} d \phi_{A},
\end{align}
which does not appear to be SO(3) covariant, but it actually is. 
This means we can that this integral is insensitive to
the rigid rotations of our coordinate axes.

\hfill \break

\begin{definition}[label=def:D]
Using the language of symplectic forms and differential geometric version 
of the generalized Stokes' theorem, we can see that 
$ \oint S_z d \phi =  \int d S_z \wedge d\phi$, the integral on the LHS is a
line integral, whereas that on the RHS is an area integral on the spin
sphere. Area integrals are indeed SO(3) covariant.
\end{definition}

\hfill \break

So, we rotate our axes so that the $z$-axis points along $\vv{n}$.
The spin sector integral is 
\begin{align}
\mathcal{J}_{A}^{\mathrm{spin}}   =\frac{1}{2 \pi} \oint S_{A}^{z} d \phi_{A} 
  & =    S_{A}^{z}   =  \hat{n} \cdot \vv{S}_A.   \label{spin_contribution}
\end{align}
We have used the fact that $S_{A}^{z}$ is constant on the loop of integration;
this is so because $\vv{S}_A$ makes a constant angle with the $z$-axis 
(or $\vv{n}$ vector) while we perform the line integral.

Finally, combining Eqs.~\eqref{orbital_contribution} and \eqref{spin_contribution},
our action integral becomes 
\begin{equation}
\mathcal{J}=\hat{n} \cdot\left(\vec{L}+\vec{S}_{1}+\vec{S}_{2}\right)=\hat{n} \cdot \vec{J} = J.
\end{equation}

\section{Computation of $\mc{J}_2$ and $\mc{J}_3$}

The procedure for computing the next two actions is very similar.
Instead of flowing under $J^2$, we flow under $L^2$ and $J_z$,
with the corresponding $\vec{n}$ being $2 \vv{L}$ and $\hat{z}$,
with the exception that under the $L^2$ flow, the spin vectors
don't move; only orbital ones do.
The amount of flow required to close the loop is still given by
$\Delta \lambda = 2 \pi/n$.

Doing similar calculations as above, we find that the 
corresponding actions turn out to be $\hat{n} \cdot \vec{J}$
which gives us 
\begin{equation}
\mathcal{J}_{2}= L , \quad \mathcal{J}_{3}=  J_z.
\end{equation}
All in all, we finally have 
\begin{equation}
\mathcal{J}_{1}= J, \quad\mathcal{J}_{2}= L , \quad \mathcal{J}_{3}=  J_z.
\end{equation} \\

\begin{tcolorbox}
From the expressions of the above three actions, we note two features
that actions appear to possess
\begin{itemize}
\item Action variables are functions of the commuting constants: $\mc{J}(\vec{C})$.
In fact, all actions are constants and are also mutually commuting. 
\item An action is a function of only those $C_i$'s under which we need to flow
to close the loop, the integral over which furnishes the action. 
\end{itemize}
\end{tcolorbox}

\section{Computation of $\mc{J}_4$}

We won't derive $\mc{J}_4$ because, this action also
has a Newtonian limit which is derived in graduate level texts; see
Eq.~(10.139) of Ref.~\cite{goldstein2013classical}. 
It's 1PN extension was worked out in Ref.~\cite{Damour:1988mr}, 
see Eq.~(3.10) therein. The 1.5PN version of this action is given 
in Eq.~(38) of Ref.~\cite{tanay2021integrability}.
The methods to achieve these PN versions of the fourth action
is similar and chiefly involves complex contour integration technique
invented by Arnold Sommerfeld.

For the reference of the reader, the fourth action is
\begin{equation}
\mathcal{J}_{4}=-L+\frac{G M \mu^{3 / 2}}{\sqrt{-2 H}}+\frac{G M}{c^{2}}\left[\frac{3 G M \mu^{2}}{L}+\frac{\sqrt{-H} \mu^{1 / 2}(\nu-15)}{\sqrt{32}}-\frac{2 G \mu^{3}}{L^{3}} \vec{S}_{\mathrm{eff}} \cdot \vec{L}\right]+\mathcal{O}\left(c^{-4}\right)
\end{equation}

\chapter{Computation of the fifth action}    \label{chapter-6}

\section{Problems in trying to compute the spin sector action integral while flowing under $\Eff$}

Under the $\Eff$ flow, we have the following
EOMs.
\begin{equation}
\begin{aligned}
&\frac{d \vec{R}}{d \lambda}=\vec{S}_{\mathrm{eff}} \times \vec{R} , \\
&\frac{d \vec{P}}{d \lambda}=\vec{S}_{\mathrm{eff}} \times \vec{P} ,  \\
& \frac{d \vec{S}_{a}}{d \lambda} =\sigma_{a}\left(\vec{L} \times \vec{S}_{a}\right),
\end{aligned}
\end{equation}
which further imply that
\begin{equation}
\begin{aligned}
\frac{d \vec{L}}{d \lambda} &=\vec{S}_{\mathrm{eff}} \times \vec{L}.
\end{aligned}
\end{equation}

We now try to compute the contribution to the action integral.
We get
\begin{equation}    \label{J5_1}
\begin{aligned}
&2 \pi \mathcal{J} =2 \pi\left(\mathcal{J}^{\mathrm{orb}}+\mathcal{J}^{\mathrm{spin}}\right) \\
&=\int_{\lambda_{i}}^{\lambda_{f}}\left(P_{i}  {d R^{i}}
+  S_1^z {d \phi_1^z}
+  S_2^z {d \phi_2^z}
\right)     \\
&=\int_{\lambda_{i}}^{\lambda_{f}}\left(P_{i} \frac{d R^{i}}{d \lambda}
+  S_1^z \frac{d \phi_1^z}{d \lambda}
+  S_2^z \frac{d \phi_2^z}{d \lambda}
\right) d \lambda
\end{aligned}
\end{equation}
Let's focus only on the orbital part.
\begin{equation}    \label{J5_orb_B}
2 \pi \mathcal{J}^{\text {orb }}=\int_{\lambda_{i}}^{\lambda_{f}} \vec{P} \cdot\left(\vec{S}_{\mathrm{eff}} \times \vec{R}\right) d \lambda=\int_{\lambda_{i}}^{\lambda_{f}}\left(S_{\mathrm{eff}} \cdot L\right) d \lambda=\left(S_{\mathrm{eff}} \cdot L\right) \Delta \lambda
\end{equation}
where we have used $S_{\mathrm{eff}} \cdot L$ to represent
$\Eff$. It could be pulled out of the integral because it is a constant 
during the $\Eff$ flow, i.e. $d{\Eff}/d \lambda =  \pb{\Eff, \Eff} = 0$.

So, the orbital sector of the integral was simple. But we have no idea how
to do the spin sector integral. Note that the simplicity of the orbital sector 
integral is owed to $\vv{L}$ being written as a cross product
$\vv{L} = \vec{R} \times \vec{P}$. The same is not true for the spins, which makes
it impossible to do the spin sector integral in Eq.~\eqref{J5_1}.

\section{Computing the spin sector  contribution to the action integral using the extended phase space}

\subsection{Introducing the fictitious variables}

The main reason the spin sector contribution to the action integral
while flowing under $\Eff$ is that spins can't be written 
a cross product of some positions and some momenta, the way $\vv{L}$
can be. This motivates us to introduce unmeasurable, fictitious variables
$\vv{R_{a}}$ and $\vv{P_{a}}$ with $a = 1, 2$ such that
\begin{equation}
\vec{S}_{a} \equiv \vec{R}_{a} \times \vec{P}_{a}  . 
\end{equation}
With these fictitious variables, the spins are no longer the 
independent, fundamental coordinates but now they rather depend on
the fictitious variables. We now introduce some simple 
terminology. 
The totality of $\vec{R}, \vec{P}, \vec{S}_1$ and $\vec{S}_2$
forms the standard phase space (SPS).
The totality of $\vec{R}, \vec{P}, \vec{R}_{1/2}$ and $\vec{P}_{1/2}$
forms the extended phase space (EPS). 
The vectors $\vec{R}_{1/2}$ and $\vec{P}_{1/2}$ form the sub-spin space.

Many things need to be put on a firm footing with the 
introduction of these new fictitious variables. We do so one by one.
\begin{itemize}
\item \textbf{Hamiltonian:} The Hamiltonian will now be seen
as a function of the EPS coordinates, rather than the SPS ones.
\item \textbf{Poisson brackets:} PBs need to be defined, for the EOMs are written
in their terms. We propose
\begin{equation}
\left\{R_{i}, P_{j}\right\} =\delta_{ij}, \quad\left\{R_{ai}, P_{b j}\right\}=\delta_{a b} \delta_{ji}      \label{EPS_PBs}
\end{equation}
\item \textbf{EOM:} The EOMs are still given by the same familiar equation
\begin{equation}
\frac{d f}{d t}=\{f, H\}
\end{equation}
\item \textbf{Equivalency of the SPS and the EPS pictures in terms of EOMs:}
The only reason we can introduce the EPS picture is that the EPS
picture is completely equivalent to the SPS one, i.e. the EOMs for all
real (non-fictitious) variables are the same in the two pictures.

Why do we say that? This is because the EPS PBs imply the SPS PBs, i.e.
Eqs.\eqref{EPS_PBs} $\implies$  Eqs.~\ref{C1-canonical PB}.
\item \textbf{Equivalency of the SPS and the EPS pictures in terms of integrability:}
The system is integrable even in the EPS picture. To show that
we need to come up with $n = 2n/2 = 18/2 = 9$ commuting constants in the EPS picture.
Five of them are the SPS commuting constants (see Eq.~\eqref{C1-CCs})
but considered functions of the EPS coordinates and not the SPS coordinates.
The next four are $S^2_{1/2}$ and $\vec{R}_{1/2} \cdot \vec{P}_{1/2}$.
\item \textbf{Sanity check 1:} All observables should depend only on non-fictitious variables alone. This is indeed the case. All five actions we derive depend only 
on $\vec{R}, \vec{P}, \vec{S}_1$ and $\vec{S}_2$ and the two masses; 
any dependence on the fictitious variables is through $\vec{S}_1$ and $\vec{S}_2$.
Although actions are not observables, frequencies $\omega_i = \pd H / \pd \mc{J}_i$
are; and they are a functions of non-fictitious variables only.
\item \textbf{Sanity check 2:} We have checked numerically that the fifth action
(computed with the help of fictitious variables), when seen as a function of 
the SPS coordinates only, generates a flow which forms a closed loop after 
flowing by $2 \pi$ within numerical errors.
\item \textbf{An EPS action is also an SPS action:} If we succeed in computing
an EPS action which is a function of ($\vec{R}, \vec{P}, \vec{S}_1$ and $\vec{S}_2$),
then this should also serve as an SPS action. How? Our EPS action flow 
must make a loop in the EPS. The same action when seen as  a function of the SPS
coordinates must also make a loop in the SPS. This is because the PBs are the 
same in the SPS and the EPS and flow equations are written using PBs. 
Since the flow of the EPS action (when seen as  a function of SPS coordinates)
makes a closed loop in the SPS, it can also be considered as a legitimate action in
the SPS.
\end{itemize}

\hfill \break

\begin{definition}[label=def:F]
There is small hole in the argument presented above regarding the EPS
action being also the SPS action. What if the EPS action loop 
(obtained by flowing under the fifth action by $2 \pi$)
corresponds to a loop in the SPS which goes around more than once.
We require actions to make a closed loop  \textit{just once}
when we flow under them and the above possibility is undesirable.
But this is not  a cause for worry because in Ref.~\cite{tanay2021action},
we invoke some topology arguments to rule this out.
\end{definition}

\hfill \break

\begin{tcolorbox}
The method of inventing temporary spurious variables is not that
uncommon. We use complex contour integration methods to compute integrals
which are cast in terms of reals and whose result is also real.
See Sec.~11.8 of Ref.~\cite{arfken2013mathematical}. 
In this case, the complex numbers can be thought of as temporary spurious
variables invented to solve some problem which was cast in terms of fewer
variables (reals).
\end{tcolorbox}

\subsection{Computing the spin sector of the action integral under the $\Eff$ flow}

In the EPS picture, under the $\Eff$ flow the EOMs are
\begin{equation}
\begin{aligned}
\frac{d \vec{R}}{d \lambda} &=\vec{S}_{\mathrm{eff}} \times \vec{R} ,  \\
\frac{d \vec{P}}{d \lambda} &=\vec{S}_{\mathrm{eff}} \times \vec{P} , \\
\frac{d \vec{R}_{a}}{d \lambda} &=\sigma_{a}\left(\vec{L} \times \vec{R}_{a}\right) ,\\
\frac{d \vec{P}_{a}}{d \lambda} &=\sigma_{a}\left(\vec{L} \times \vec{P}_{a}\right).
\end{aligned}
\end{equation}
The contribution to the action integral becomes
\begin{equation}
\mathcal{J}_k=\frac{1}{2 \pi} \oint_{\mathcal{C}_{k}}\left(\vec{P} \cdot d \vec{R}+\vec{P}_{1} \cdot d \vec{R}_{1}+\vec{P}_{2} \cdot d \vec{R}_{2}\right),
\end{equation}
which further entails that
\begin{align}
&2 \pi \mathcal{J}_{S_{\text {eff }} \cdot L}=2 \pi\left(\mathcal{J}^{\text {orb }}+\mathcal{J}^{\text {spin }}\right)\\
&=\int_{\lambda_{i}}^{\lambda_{f}}\left(P_{i} \frac{d R^{i}}{d \lambda}+P_{1 i} \frac{d R_{1}^{i}}{d \lambda}+P_{2 i} \frac{d R_{2}^{i}}{d \lambda}\right) d \lambda\\
&=\int_{\lambda_{i}}^{\lambda_{f}}\left(\vec{P} \cdot\left(\vec{S}_{\mathrm{eff}} \times \vec{R}\right)+\vec{P}_{1} \cdot\left(\sigma_{1} \vec{L} \times \vec{R}_{1}\right)\right.\\
&\left.+\vec{P}_{2} \cdot\left(\sigma_{2} \vec{L} \times \vec{R}_{2}\right)\right) d \lambda\\
&=2 \int_{\lambda_{i}}^{\lambda_{f}}\left(S_{\mathrm{eff}} \cdot L\right) d \lambda=2\left(S_{\mathrm{eff}} \cdot L\right) \Delta \lambda_{S_{\mathrm{eff}} \cdot L}\\
&\mathcal{J}_{S_{\mathrm{eff}} \cdot L}=\frac{\left(S_{\mathrm{eff}} \cdot L\right) \Delta \lambda_{S_{\mathrm{eff}} \cdot L}}{\pi},   \label{J5_seffdl}
\end{align}
which is just twice the contribution of 
the orbital part given in Eq.~\eqref{J5_orb_B}.

\begin{tcolorbox}
Basically, with the introduction of the fictitious variables, and expanding the 
SPS to EPS, we bring the spins at the same mathematical footing as $\vec{L}$, in that
all three of them can be written as cross products of some position with some 
momentum. This renders the otherwise insoluble spin sector contribution
to the action integral under the $\Eff$ flow rather trivial to deal with (the way
orbital sector contribution is), as if it was meant to be.
\end{tcolorbox}

\section{Computing the fifth action}

\subsection{Setting up the stage}

We will only outline the steps needed to compute the fifth action. For details,
refer to Ref.~\cite{tanay2021action}. To compute the fifth action,
a flow only under $\Eff$ is not enough; it won't make a closed loop.
In the SPS space, to close the loop, we further need to flow under
$J^2$ and $L^2$, whereas in the EPS space, we need two more flows 
($S_1^2$ and $S_2^2$ flow) on the
top of that because we have extra variables in the EPS. Because we know
how to compute the action integral  contribution corresponding
to $\Eff$ flow only in the EPS, we will try to close the loop in the EPS.
The successive flows we need are those of $\Eff, J^2, L^2, S_1^2$ and $S_2^2$.
At this point, we mention the result that under the flow by 
the last four 
constants of motion, the contribution to the action integral is 
\begin{equation}
\begin{aligned}
&\mathcal{J}_{J^{2}}=\frac{J^{2} \Delta \lambda_{J^{2}}}{\pi} , \\
&\mathcal{J}_{L^{2}}=\frac{L^{2} \Delta \lambda_{L^{2}}}{\pi}, \\
&\mathcal{J}_{S_{1}^{2}}=\frac{S_{1}^{2} \Delta \lambda_{S_{1}^{2}}}{\pi}, \\
&\mathcal{J}_{S_{2}^{2}}=\frac{S_{2}^{2} \Delta \lambda_{S_{2}^{2}}}{\pi}.
\end{aligned}
\end{equation}
These are easy to derive and can be derived in an
analogous manner as the contribution corresponding to the first
constant $\Eff$  (see Eq.~\eqref{J5_seffdl}). Therefore the 
fifth action becomes
\begin{equation}     \label{J5_final}
\begin{aligned}
\mathcal{J}_{5}=& \frac{1}{\pi}\left\{\left(S_{\mathrm{eff}} \cdot L\right) \Delta \lambda_{S_{\mathrm{eff}} \cdot L}+J^{2} \Delta \lambda_{J^{2}}+L^{2} \Delta \lambda_{L^{2}}
+S_{1}^{2} \Delta \lambda_{S_{1}^{2}}+S_{2}^{2} \Delta \lambda_{S_{2}^{2}}\right\} .
\end{aligned}
\end{equation}
The problem of determining the fifth action thus reduces to determining 
the flow amounts $\Delta \lambda$'s needed (under various commuting constants) to close the loop.

\begin{figure}
  \centering
  \includegraphics[width=0.8\linewidth]{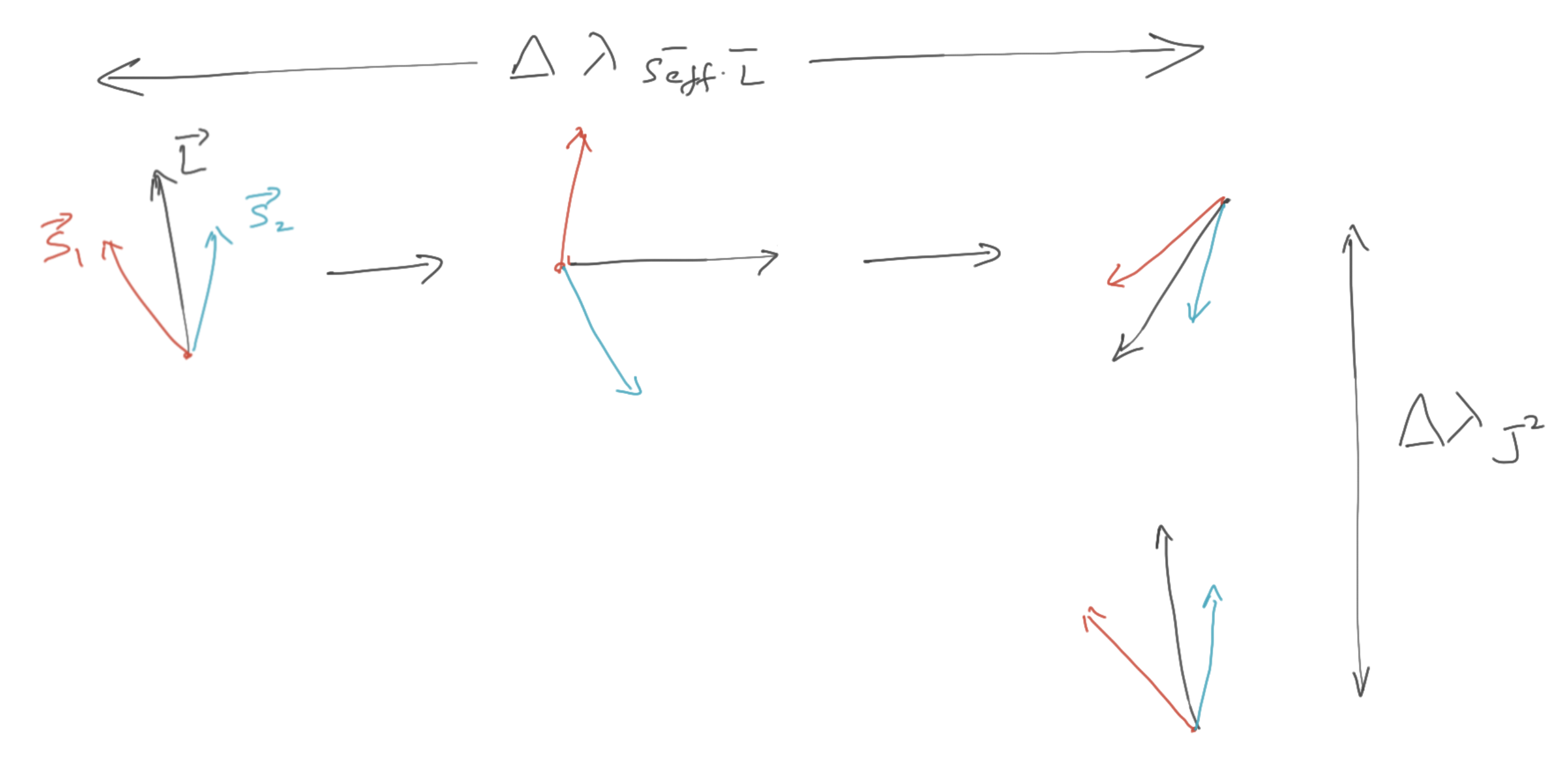}
  \caption{ Behavior of the angular momenta under 
  the $\Eff$ flow.
    \vspace{-1.em}
  }
  \label{seffdl_flow}
\end{figure}

\subsection{Determining the flow amounts}

Please see Ref.~\cite{tanay2021action} for details on how to determine the flow
amounts under various commuting constants. Here we briefly sketch the overall picture.

\subsubsection{$\Eff$ flow}

Under the $\Eff$ flow the magnitudes of all the three 3D position and 
momentum vectors stay constant. So does the magnitude of all the three
angular momenta. The effect of $\Eff$ flow on the three angular momenta 
is shown in Fig.~\ref{seffdl_flow}. The triad formed by the three
angular momenta acts like a lung and it ``inhales'' and ``exhales''.
All three mutual angles between the angular momenta are periodic functions
of the flow parameter with the same period $\Delta \lambda_{\Eff}$.
See Ref.~\cite{tanay2021action}
for the derivation of these results. So, naturally we want to flow 
under $\Eff$ by exactly by a multiple of this period because if we 
did not, then there is very little hope of restoring all the coordinates
to their initial state by flowing under other constants. This
is so because other constants do not change these mutual angles
between the three angular momenta (except for the Hamiltonian).
So, we decide to flow under $\Eff$ by an amount $\Delta \lambda_{\Eff}$.

\subsubsection{$J^2$ flow}

It's shown in Ref.~\cite{tanay2021action} that by an appropriate amount
$\Delta \lambda_{J^2}$ of flow under $J^2$, we can restore all the three angular momenta to their initial states.

\subsubsection{$L^2, S_1^2, S_2^2$ flow}

Because all three angular momenta have been restored, the only
way $\vec{R}, \vec{P}, \vec{R}_{1/2}$ and $\vec{P}_{1/2}$ off from their 
initial state is by a some finite rotation in the plane perpendicular
to $\vec{L}$ (for $\vec{R}, \vec{P}$) 
and $\vec{S}_{1/2}$ (for $\vec{R}_{1/2}, \vec{P}_{1/2}$). To negate this offset,
all we need to do is to flow under $L^2, S_1^2$ and $S_2^2$ by some
appropriate amounts. This is also worked out in Ref.~\cite{tanay2021action}.

Once these amounts are determined Eq.~\eqref{J5_final}
finally yields the fifth action variable, where the $\Delta \lambda$'s are 
given by Eqs.~(A42), (A67), (A77), (A94) and (A95) of Ref.~\cite{tanay2021action}.

\bibliographystyle{unsrt}
\bibliography{pn_v2}
\end{document}